\documentclass[twocolumn,trackchanges]{aastex62}

\let\pwiflocal=\iffalse \let\pwifjournal=\iffalse
\usepackage[utf8]{inputenc}
\usepackage{textcomp}
\usepackage{amsmath,amssymb,gensymb,times,graphicx,morefloats,color}
\usepackage{afterpage, bm}
\usepackage{natbib,hyperref}
\usepackage[outdir=./]{epstopdf}
\usepackage{mathrsfs}

\newcommand{\um}{$\mu$m}
\newcommand{\fbol}{$F_{\mathrm{bol}}$}

\newcommand{\ms}{m~s$^{-1}$}

\def\spitzer{\emph{Spitzer}}
\def\Spitzer{\emph{Spitzer}}

\def\tess{\emph{TESS}}

\newcommand{\targ}{HIP~67522}
\newcommand{\ticname}{TIC~166527623}

\shorttitle{A Transiting Planet in the Sco-Cen Association}
\shortauthors{Rizzuto et al.}

\begin{document}

\title{TESS Hunt for Young and Maturing Exoplanets (THYME) II:
A 17\,Myr Old Transiting Hot Jupiter in the Sco-Cen Association}
\correspondingauthor{Aaron C. Rizzuto} 
\email{arizz@astro.as.utexas.edu}

\author[0000-0001-9982-1332]{Aaron C. Rizzuto}
\altaffiliation{51 Pegasi b Fellow}
\affiliation{Department of Astronomy, The University of Texas at Austin, Austin, TX 78712, USA}

\author[0000-0003-4150-841X]{Elisabeth R. Newton}
\affiliation{Department of Physics and Astronomy, Dartmouth College, Hanover, NH 03755, USA}

\author[0000-0003-3654-1602]{Andrew W. Mann}
\affiliation{Department of Physics and Astronomy, The University of North Carolina at Chapel Hill, Chapel Hill, NC 27599, USA} 

\author[0000-0003-2053-0749]{Benjamin M. Tofflemire}
\altaffiliation{51 Pegasi b Fellow}
\affiliation{Department of Astronomy, The University of Texas at Austin, Austin, TX 78712, USA}

\author[0000-0001-7246-5438]{Andrew Vanderburg}
\altaffiliation{NASA Sagan Fellow}
\affiliation{Department of Astronomy, The University of Texas at Austin, Austin, TX 78712, USA}

\author[0000-0001-9811-568X]{Adam L. Kraus}

\author[0000-0001-7336-7725]{Mackenna L. Wood}
\affiliation{Department of Physics and Astronomy, The University of North Carolina at Chapel Hill, Chapel Hill, NC 27599, USA} 

\author[0000-0002-8964-8377]{Samuel N. Quinn}
\affiliation{Center for Astrophysics, Harvard \& Smithsonian, 60 Garden St, Cambridge, MA 02138, USA}

\author[0000-0002-4891-3517]{George Zhou}
\affiliation{Center for Astrophysics, Harvard \& Smithsonian, 60 Garden St, Cambridge, MA 02138, USA}
\altaffiliation{Hubble Fellow}

\author[0000-0001-5729-6576]{Pa Chia Thao}
\altaffiliation{NSF GRFP Fellow}
\affiliation{Department of Physics and Astronomy, The University of North Carolina at Chapel Hill, Chapel Hill, NC 27599, USA}

\author[0000-0001-9380-6457]{Nicholas M. Law}
\affiliation{Department of Physics and Astronomy, The University of North Carolina at Chapel Hill, Chapel Hill, NC 27599, USA}

\author[0000-0002-0619-7639]{Carl Ziegler}
\affiliation{Dunlap Institute for Astronomy and Astrophysics, University of Toronto, 50 St. George Street, Toronto, Ontario M5S 3H4, Canada}

\author[0000-0001-7124-4094]{C\'esar Brice\~no}
\affiliation{Cerro Tololo Inter-American Observatory, Casilla 603, La Serena, Chile}





\begin{abstract}

        We present the discovery of a transiting hot Jupiter orbiting \targ~ ($T_{eff}\sim5650$ K; $M_* \sim 1.2 M_{\odot}$) in the 10-20\,Myr old Sco-Cen OB association. We identified the transits in the \tess{} data using our custom notch-filter planet search pipeline, and characterize the system with additional photometry from \spitzer{}, spectroscopy from SOAR/Goodman, SALT/HRS, LCOGT/NRES, and SMARTS/CHIRON, and speckle imaging from SOAR/HRCam. We model the photometry as a periodic Gaussian process with transits to account for stellar variability, and find an orbital period of 6.9596$^{+0.000016}_{-0.000015}$\,days and radius of 10.02$^{+0.54}_{-0.53}$\,R$_\oplus$. We also identify a single transit of an additional candidate planet with radius 8.01$^{+0.75}_{-0.71}$\,R$_\oplus$ that has an orbital period of $\gtrsim23$\,days. The validated planet HIP\,67522\,b is currently the youngest transiting hot Jupiter discovered and is an ideal candidate for transmission spectroscopy and radial velocity follow-up studies, while also demonstrating that some young giant planets either form in situ at small orbital radii, or else migrate promptly from formation sites farther out in the disk.
    
\end{abstract}

\keywords{exoplanets, exoplanet evolution, young star clusters- moving clusters, planets and satellites: individual (HIP 67522)}


\section{Introduction}
The properties of exoplanets are likely a strong function of their age, reflecting the formation and evolutionary processes that shape them. Dynamical interactions and migration, whether through interaction with a circumstellar disk \citep{lubow_migration}, or other planets in the system (e.g., \citealt{ChatterjeeDynamical2008,fabrycky07}) can alter orbital properties, photoeveporation from high energy radiation or core-heating can impact atmospheric properties (e.g., \citealt{Ehrenreich2015,gupta19}), and the details of the greater star cluster environment can sculpt system architectures \citep{cai17,elteren19}. The conditions and timescales under which each of these various mechanisms dominate is currently unclear, though the majority of evolution is expected to occur in the first few hundred million years following formation \citep{Adams2006,Mann2010,Lopez2012}.  Most observations of exoplanets are generally of mature, $>$1\,Gyr old systems, and as such only directly reflect the end stage population.

Detection and characterization of young exoplanets provides one of the most direct avenues to mapping exoplanet evolution, and observations of exoplanets in clusters or associations of known age are particularly useful because they inherit the bulk properties such as age and metallicity of the host clusters to a level of precision not normally measurable for single field stars. Young exoplanets have been detected as part of direct imaging surveys (e.g., \citealt{bowler_review,NielsenGPIES}), with radial velocities in young open clusters (e.g., \citealt{quinn14}), and with transits observed by the repurposed \emph{Kepler} mission (\emph{K2}) \citep{Howell2014} which observed several open clusters and associations (e.g., \citealt{zeit1,trevor_k233,zeit4,zeit8,trevorv1298}).

The Transiting Exoplanet Survey Satellite (\tess{}, \citealt{rickertess_2014}) mission provides the first opportunity for large-scale transit searches in young associations and young moving groups with ages $<$300\,Myr.  In particular, during the first year of observations, \tess{} observed the majority of the Scorpius-Centaurus OB association (Sco-Cen), a 10-20\,Myr old, low density comoving population of some $\sim$10000 members \citep{Pecaut2012,wifes1_2015}. Sco-Cen is the nearest region of recently completed star formation to the sun ($d~\sim140$\,pc; \citealt{zeeuw99}), and contains the vast majority of pre-main-sequence stars within the nearest 200\,pc.

In this paper, we present the discovery of a hot, Jupiter-sized transiting exoplanet in the Sco-Cen association in the \tess{} Sector 11 dataset. The host star is \targ{}, an early G-Type member of the Sco-Cen association originally identified kinematically by \citet{zeeuw99}. We also detect a single transit of a second companion in the system, with a radius that is closer to that of Saturn and an orbital period likely beyond the time baseline of a \tess{} observing sector. In Section \ref{sec:obs} we present the \tess{} photometry, follow-up \spitzer{} transit photometry, and other ground-based followup. In Section \ref{sec:measure} we derive stellar parameters, radial velocities, and analyse the rotation of the host star. In Section \ref{sec:membership} we reassess the membership of \targ~ in Sco-Cen in light of recent astrometric and photometric catalogs, and in Section \ref{sec:tfits} we present the results of transit model fits to the lightcurves from both \tess{} and \spitzer{} and address false-positive scenarios. Finally, in Section \ref{sec:discussion} we discuss the system architecture, the implications for migration and formation of gas-giant exoplanets and the potential for follow-up characterization.

\begin{deluxetable*}{ccc}
\centering
\tabletypesize{\scriptsize}
\tablewidth{0pt}
\tablecaption{Properties of the host star \targ~ (\ticname). \label{proptab}}
\tablehead{\colhead{Parameter} & \colhead{Value} & \colhead{Source} }
\startdata
\multicolumn{3}{c}{Astrometry}\\
\hline
R.\,A.  & 13 50 06.280 & TIC\\
Decl. & -40 50 08.88 & TIC\\
$\mu_\alpha$ (mas\,yr$^{-1}$)& -28.843$\pm$0.108 & \emph{Gaia} DR2\\
$\mu_\delta$  (mas\,yr$^{-1}$) & -22.425$\pm$0.107 & \emph{Gaia} DR2\\
$\pi$ (mas) & 7.8288$\pm$0.0671 & \emph{Gaia} DR2\\
\hline
\multicolumn{3}{c}{Photometry}\\
\hline
G$_{Gaia}$ (mag) & 9.6326$\pm$0.0023 & \emph{Gaia} DR2\\
BP$_{Gaia}$ (mag) & 9.9859$\pm$0.0038 & \emph{Gaia} DR2\\
RP$_{Gaia}$ (mag) & 9.1359$\pm$0.0038 & \emph{Gaia} DR2\\
B (mag) & 10.571$\pm$0.014 & APASS\\
B$_T$ (mag) & 10.591$\pm$0.036 & TYCHO 2 \\
V$_T$ (mag) & 9.876$\pm$0.026 & TYCHO 2\\
g' (mag) & 10.830$\pm$0.030  & APASS\\
J (mag) &8.587$\pm$0.021& 2MASS\\
H (mag) &8.287$\pm$0.042& 2MASS\\	
Ks (mag) &8.164$\pm$0.026& 2MASS\\
W1 (mag) & 8.108$\pm$0.022 & ALLWISE\\
W2 (mag)& 8.122$\pm$0.019 & ALLWISE\\
W3 (mag)& 8.077$\pm$0.018 & ALLWISE\\ 
W4 (mag)& 8.121$\pm$0.186 & ALLWISE\\ 
\hline
\multicolumn{3}{c}{Kinematics \& Position}\\
\hline
Barycentric RV (km\, s$^{-1}$) & 7.41$\pm$0.25 & This paper\\
U (km\, s$^{-1}$) &7.12$\pm$0.19 & This paper\\
V (km\, s$^{-1}$) & -21.53$\pm$0.22 & This paper\\
W (km\, s$^{-1}$) & -5.67$\pm$0.13 & This paper\\
X (pc) & -84.15$\pm$0.73 & This paper\\
Y (pc) & -84.89$\pm$0.74 & This paper\\
Z (pc) & 45.13$\pm$0.39 & This paper\\
Distance (pc) &127.27$^{+1.10}_{-1.08}$& \citet{gaia_distances} \\
\hline
\multicolumn{3}{c}{Physical Properties}\\
\hline
Rotation Period (days) &1.418$\pm$0.016 & This paper\\
$v\sin{i}$ (km\, s$^{-1}$) & 54.2$\pm$0.7 & This paper\\
\fbol\,(erg\,cm$^{-2}$\,s$^{-1}$)& $(3.44\pm0.17)\times10^{-9}$ & This paper\\
T$_{\mathrm{eff}}$ (K) &5675$\pm$75& This paper\\
M$_\star$ (M$_\odot$) & 1.22$\pm$0.05 & This paper \\
R$_\star$ (R$_\odot$) &  1.38$\pm$0.06 & This paper \\
L$_\star$ (L$_\odot$) & 1.75$\pm$0.09 & This paper \\
$\rho_\star$ ($\rho_\odot$) & 0.46$\pm$0.06 & This paper \\
Age (Myr) & 17$\pm$2 & This paper \\
E(B-V) (mag) &  0.04$\pm$0.02 & This paper\\
\enddata
\end{deluxetable*}

\section{Observations}\label{sec:obs}

\subsection{TESS Photometry} \label{sec:tessphot}
\tess{} observed \targ~during the eleventh sector of observations between 2019 April 22 and 2019 May 21. \targ~was pre-selected for short cadence (two-minute) observations on the basis of its membership in the Sco-Cen association  as part of our GI proposal G011280 (PI Rizzuto). The data were processed by the Science Processing and Operations Center (SPOC) pipeline at NASA Ames \citep{Jenkins:2015,Jenkins:2016}, which performed pixel calibration, extracted lightcurves, deblended light from nearby objects, removed common-mode systematics, high pass filtered the lightcurve, and searched for transits. Due to the high amplitude rotational variability produced by spots ($\sim$1-2\%; $P \sim 1.4$ days) which featured rotational structure on $\lesssim$1\,day timescales, the SPOC transit search pipeline registered a threshold crossing event consisted with our detection, but rejected it as non-planetary.

We applied the detrending and transit-search algorithm of \citet{zeit5} to the SPOC lightcurve of \targ, with a 0.5\,day filtering window and transit durations of up to 5\,hours. This method removes astrophysical variability with a notch filter, which fits a window of the lightcurve as a combination of an outlier-robust second-order polynomial and a trapezoidal notch. The inclusion of the notch allows detrending outside the notch without removal or over correction of transit-like signals. The window is then moved along each point in the lightcurve, detrending variability signals from the entire dataset. In the case of \targ, the rapid variability was not completely removed, as there is degeneracy between real transits and rotational signals of similar timescale. We improved the detection by taking into account the shape of potential transits with the notch filter. At each data point, we calculate the difference in Bayesian Information Criterion (BIC) for a model consisting of only the polynomial, and a model including the notch. This difference will then be larger for transit-like shapes with rapid ingress and egress than for spot-induced variability on transit timescales. We searched for periodic transits in the timeseries of this BIC observable using the Box Least Squares (BLS) algorithm \citep{kovacsBLS}.  A 13.92\,day periodic signal was identifiend in the BLS power spectrum, corresponding to transits of approximately 0.5\% depth and a distinctly transit-like shape (\targ\,b). Visual inspection of the two candidate transits confirms that their shape is inconsistent with any other rotational variability profiles in the lightcurve of \targ. We also identify an additional single transit of a second planet candidate in the lightcurve (see Section \ref{sec:singletrans}). 

\begin{figure*}
    \centering
    \includegraphics[width=\textwidth]{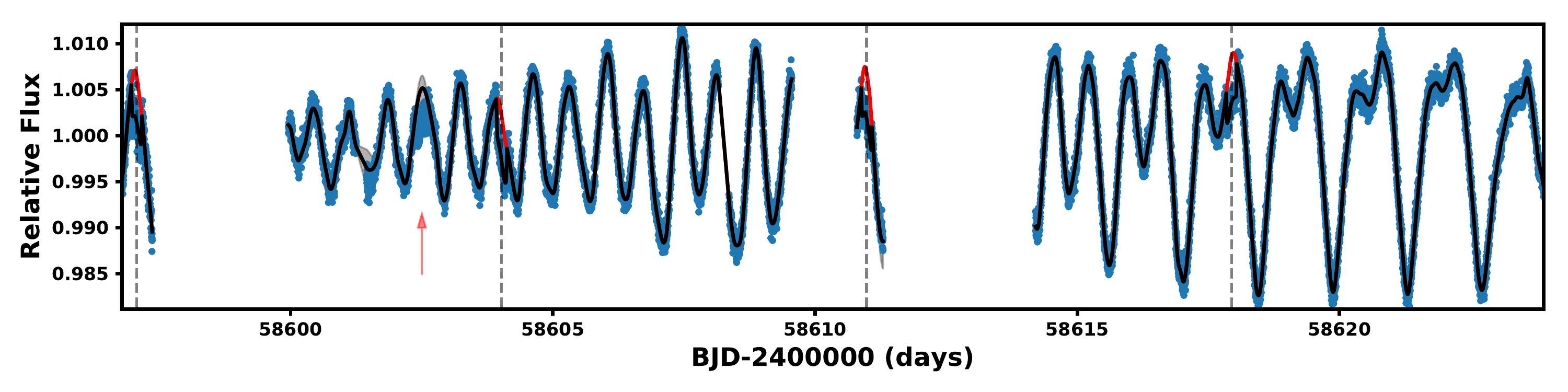}\\
    \caption{The lightcurve of \targ~ from \tess{} sector 11, corrected for systematics using the cotrending basis vector method described in \citet{smith_cbv_2012} (blue points), and the Gaussian Process model from the simultaneous transit and variability fit described in Section \ref{sec:transfit} (solid black line). Red sections showing the out-of-transit model and marked by vertical dashed lines indicate the four transits of the $\sim$7 day planet. Gray shaded regions indicate the uncertainty in the stellar rotation model, and correspond to masked regions. The masked single transit of the \targ\,c candidate is marked by a red arrow.}
    \label{fig:LCplot}
\end{figure*}

The \tess{} sector 11 data from the SPOC pipeline features a large gap between the two \tess{} orbits caused by the occurrence of a large scattered light event. We re-inspected the original postage stamps of this section of data, and a similar discarded section at the beginning of sector 11. We removed instrumental systematics following the quaternion methodology of \citet{vanderburg2019} and identify two additional transits of \targ\,b, revealing the true period to be 6.96\,days. Figure \ref{fig:LCplot} shows the SPOC lightcurve with the additional sections, as well as the epochs of the transits.

\subsection{Spitzer IRAC Photometry}\label{sect:spitzerphot}
Following the detection of transits with \tess{}, we scheduled observations of a transit of HIP~67522\,b with the \Spitzer{} Space Telescope. The observations were taken on 2019 December 09 UTC (program ID 14011, PI Newton) using the Infrared Array Camera in channel 2 at 4.5\,$\mu$m (IRAC; \citealt{FazioInfrared2004}). We took a total of 19968 frames in the 32x32 subarray, each of 1.9\,seconds, spanning a total of 11.2\,hours. As suggested by \citet{IngallsIntrapixel2012,IngallsRepeatability2016}, we placed the target on the location of peak sensitivity in IRAC channel 2, and used the ``peak-up'' pointing mode to keep the position of the star fixed to within 0.5\,pixels using the \Spitzer{} Pointing Calibration Reference Sensor. 

We processed the \spitzer{} images to produce lightcurves using the Photometry for Orbits, Eccentricities, and Transits pipeline (POET\footnote{\url{https://github.com/kevin218/POET}}, \citealt{StevensonTransit2012,Campo_2011}). This process included flagging and masking of bad pixels and calculation of Barycentric Julian Dates for each frame. The centroid position of the star in each image was then estimated by fitting a 2-dimensional, elliptical Gaussian to the PSF in a 15 pixel square centered on the target position \citep{stevenson2010}. Raw photometry was produced using simple aperture photometry using apertures of 3.0, 3.25, 3.5, 3.75 and 4.0 pixel diameters, each with a sky annulus of 9 to 15 pixels. Upon inspection of the resulting raw lightcurves, we saw no significant difference based on the choice of aperture size, and so the 3.5\,pixel aperture was used for the rest of the analysis. 

\spitzer{} detectors have significant intra-pixel sensitivity variations that can produce time dependent variability in the photometry of a target star as the centroid position of the star image moves on the detector \citep{IngallsIntrapixel2012}. To correct for this source of potential systematics, we applied the BiLinearly Interpolated Subpixel Sensitivity (BLISS) Mapping technique of \citet{StevensonTransit2012}, which was provided as an optional part of the POET pipeline. BLISS fits a model to the transit of an exoplanet that consists of a series of time-dependent functions, including a ramp and a transit model, and a spatially dependent model that maps sensitivity to centroid position on the detector. There are several choices of ramp models that can be used to model the out-of-transit variability, and usually a linear or quadratic is used for IRAC channel 2. We found that there was significant variability in the lightcurve of HIP~67522 that was not well fit by those standard choices of model. This is most likely due to the presence of both a detector ramp and stellar rotational variability in the lightcurve. We instead found that a sine function matched the data well. We modelled the planet transit as a symmetric eclipse without limb-darkening, as the purpose of this model was to remove time-dependent systematics and leave a spatially dependent sensitivity map. The time-dependent component of the model consisted of the mid-transit time (T$_0$), transit duration (T$_{14}$), ingress and egress times (T$_{12/34}$), planet to star radius ratio ($R_P/R_\star$), system flux, ramp period, ramp phase, ramp amplitude, and ramp constant offset. These parameters were explored with an Markov Chain Monte Carlo (MCMC) process, using 4 walkers with 100000 steps and a burn in region of 50000 steps. At each step, the BLISS map was computed after subtraction of the time dependent model components. 

\begin{figure}
    \centering
    \includegraphics[width=0.4\textwidth]{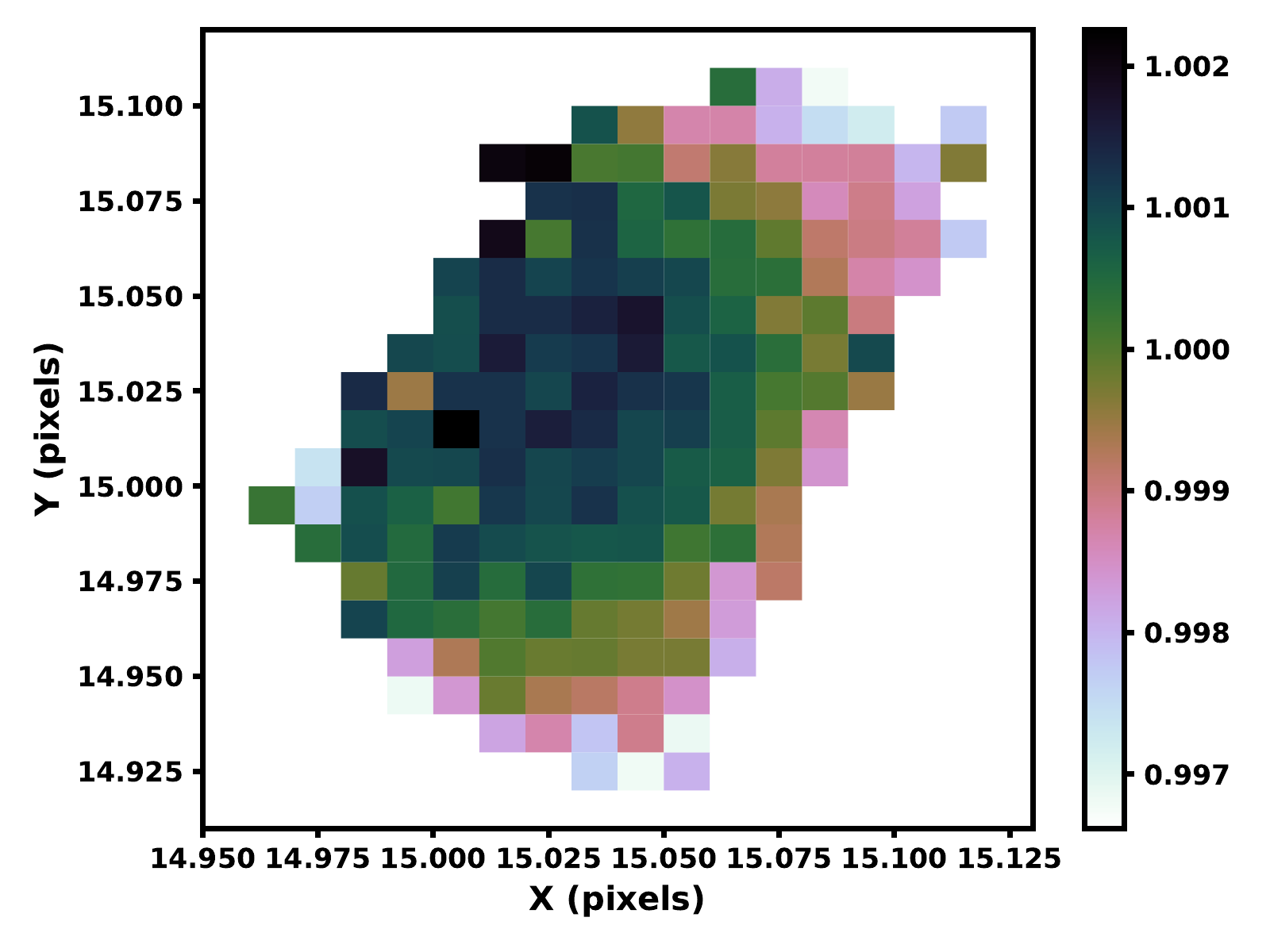}
    \caption{BLISS map of the relative sub-pixel sensitivity.}
    \label{fig:blissmap}
\end{figure}

\begin{figure}
    \centering
    \includegraphics[width=0.4\textwidth]{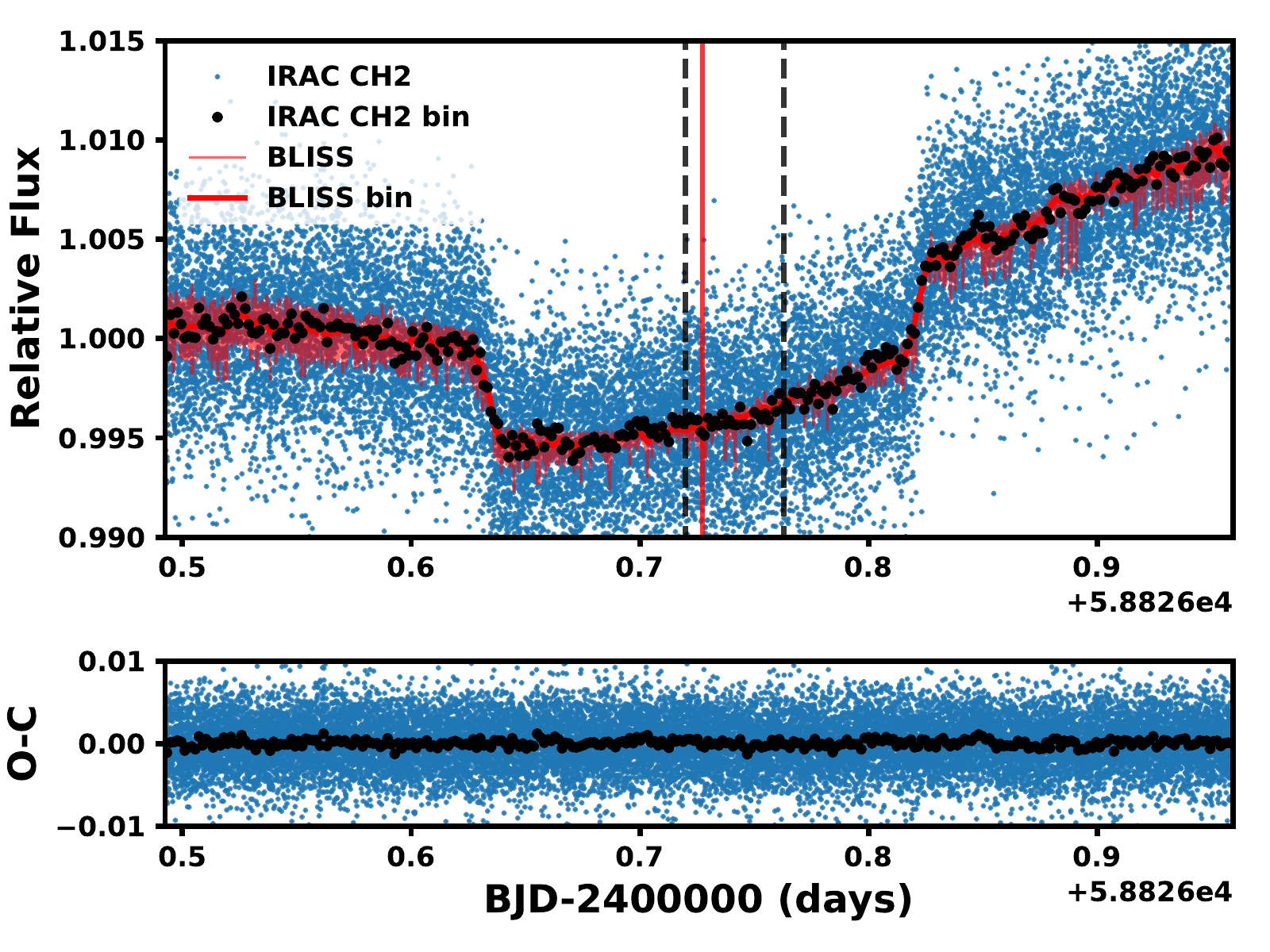}
    \caption{\emph{Upper:} \spitzer{} IRAC channel 2 data of a transit of HIP~67522\,b. Blue points are the raw Spitzer lightcurve, black circles are the same data binned into 70 bins, and the red line is the best fit BLISS model consisting of a sine ramp, transit, and subpixel sensitivity map. The red vertical line indicates the best fit center of transit, and the vertical black dashed lines indicate the 2-sigma range of the expected transit center based on the \tess{} detection. \emph{Lower:} The residuals after subtraction of the best fit model.}
    \label{fig:blissmodelplot}
\end{figure}

Table \ref{tab:blissfit} lists the best fit parameters, Figure \ref{fig:blissmap} shows the intra-pixel sensitivity BLISS map, and Figure \ref{fig:blissmodelplot} shows the \spitzer{} lightcurve of HIP~67522\,b, the best fit BLISS model and residuals, and the expected position of the transit center based on the \tess{} detection. We find that the center of the transit in the \Spitzer{} data is within 2-sigma of the expected position based on our model of the TESS lightcurve data (see Section \ref{sec:transfit}). We then subtract the spatial component of the BLISS model, namely the subpixel sensitivity map, yielding a lightcurve corrected for positional systematics, which we use in a combined transit fit with the \tess{} data (Section \ref{sec:transfit}).

\begin{deluxetable}{cc}
\tabletypesize{\scriptsize}
\tablewidth{0pt}
\tablecaption{Spitzer IRAC CH2 BLISS model fit parameters for HIP~67522\,b. \label{tab:blissfit}}
\tablehead{\colhead{Parameter} & \colhead{HIP~67522\,b}}
\startdata
T$_0$ (BJD)                          & 2458826.72742$^{+0.00028}_{-0.00027}$             \\ 
T$_{14}$ (days)                        & 0.01444$\pm$0.00011                   \\ 
$R_P/R_\star$                   & 0.0683$\pm$0.0011 \\
T$_{12/34}$ (days)                    & 0.00097$\pm$0.00011                          \\ 
System Flux ($\mu Jy$)         & 88208$^{+2271}_{-2198}$                           \\ 
Sine Amplitude                 & 14.4$^{+3.9}_{-4.3}$                             \\ 
Sine Period (days)             & 3.29$^{+0.47}_{-0.50}$                            \\ 
Sine Phase Offset              & 0.17$^{+0.13}_{-0.12}$                            \\ 
Constant                       & 15.5$^{+4.3}_{-3.9}$                              \\ 
\tess{} T$_0^a$ (BJD)                     & 2458826.741 $\pm$ 0.0152                          \\ 
\enddata
\tablenotetext{a}{Taken from the transit fit to only the \tess~mission lightcurve in Table \ref{tab:transfit}}
\end{deluxetable}

\subsection{Limits on Companions from Imaging and Gaia} 
\begin{figure}
    \centering
    \includegraphics[width=0.5\textwidth]{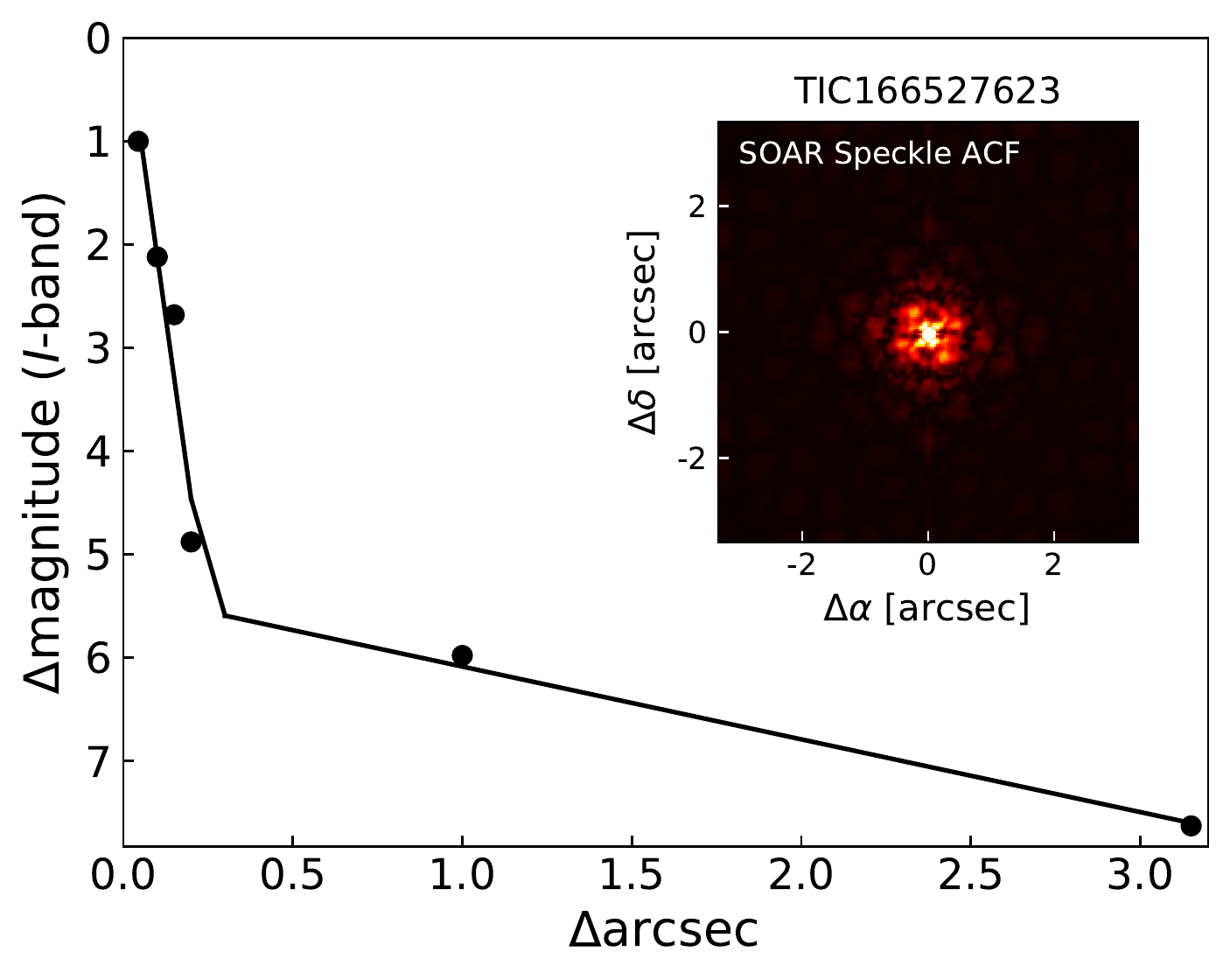}
    \caption{Detection limits from SOAR speckle imaging for \targ.}
    \label{fig:speckle}
\end{figure}

To rule out any unresolved companions that might impact our interpretation of the transit, we obtained speckle imaging using the High-Resolution  Camera (HRCam) at the 4.1m Southern Astrophysical  Research (SOAR) telescope located in Cerro Pach\'{o}n, Chile. We obtained four datacubes, two with 2$\times$2 binning and two without binning. All images were taken in the $I$-band, and reduced according to \citet{soar_tess_paper}. We show the contrast limits in Figure~\ref{fig:speckle}. The SOAR speckle imaging has an outer working angle of $\sim3$ arcseconds. No companion was detected out to the outer working angle of SOAR speckle (3\arcsec); wider companions than this should be detected by Gaia and included in the second data release \citep{GaiaDr2}.

\begin{table}
\centering
\tabletypesize{\scriptsize}
\tablewidth{0pt}
\caption{Combined contrast limits for companions to HIP~67522 from SOAR speckle imaging and Gaia astrometry. Corresponding mass limits are taken from the BHAC15 isochrone of 17\,Myr age \citep{bhac15}. \label{tab:contrastlims}}
\begin{tabular}{lccccccc}
\hline
\multicolumn{8}{c}{SOAR Speckle}\\
\hline
 $\rho$     ('')& 0.05 & 0.1 & 0.2 & 0.25 & 1.0 & 3.1 & \\
 $\rho$     (AU)& 6.4 & 13 & 26 & 32   & 128 & 396 & \\
$\Delta$I$^a$ (mag) & 1.0 & 2.2 & 2.7 & 4.8  & 5.9 & 7.5 & \\
Mass (M$_\odot$)& 1.0 & 0.66 & 0.51 & 0.16 & 0.08 & 0.03 \\
\hline
\multicolumn{8}{c}{Gaia DR2 Astrometry}\\
\hline
$\rho$    ('') & 0.03 & 0.08 & 0.20 & 2 & 3 & 6  & $>$7\\
$\rho$    (AU) & 4 & 10 & 25 & 255 & 383 & 766 & $>$900\\ 
$\Delta$G (mag) & 0    & 4    & 5    & 6 & 8 & 10 & 11.0\\
Mass (M$_\odot$)& 1.2. & 0.5  & 0.3. & 0.17  & 0.06 &0.02 & 0.015\\
\hline
\end{tabular}
\end{table}

Our null detection from speckle interferometry is consistent with the deeper limits set by the lack of {\it Gaia} excess noise, as indicated by the Renormalized Unit Weight Error \citep{GaiaDr2}\footnote{\url{https://gea.esac.esa.int/archive/documentation/GDR2/Gaia_archive/chap_datamodel/sec_dm_main_tables/ssec_dm_ruwe.html}}. HIP 67522 has $RUWE = 0.91$, consistent with the distribution of values seen for single stars. Based on a calibration of the companion parameter space that would induce excess noise (\citealt{zeit8}; Kraus et al., in preparation), this corresponds to contrast limits of $\Delta G \sim 0$ mag at $\rho = 30$ mas, $\Delta G \sim 4$ mag at $\rho = 80$ mas, and $\Delta G \sim 5$ mag at $\rho \ge 200$ mas. The evolutionary models of \citet{bhac15} would imply corresponding physical limits for equal-mass companions at $\rho \sim 4$ AU, $M \sim 0.5 M_{\odot}$ at $\rho \sim 10$ AU, and $M \sim 0.3 M_{\odot}$ at $\rho > 25$ AU. 

The Gaia DR2 catalog \citep{GaiaDr2} also does not report any comoving, codistant neighbors within $\rho < 300\arcmin$ ($\rho < 38,000$ AU) from HIP 67522. At separations beyond this limit, any neighbor would be more likely to be an unbound member of Sco-Cen, rather than a bound binary companion (e.g., \citealt{kraus08}), so we conclude that there are no wide companions to HIP 67522 above the Gaia catalog's completeness limit. \citet{zeigler18} and \citet{brandeker19} have mapped the completeness limit close to bright stars to be $\Delta G \sim 6$ mag at $\rho = 2\arcsec$, $\Delta G \sim 8$ mag at $\rho = 3\arcsec$, and $\Delta G \sim 10$ mag at $\rho = 6\arcsec$. The evolutionary models of \citet{bhac15} would imply corresponding physical limits of $M \sim 0.15 M_{\odot}$ at $\rho = 255$ AU, $M \sim 0.06 M_{\odot}$ at $\rho = 383$ AU, and $M \sim 0.02 M_{\odot}$ at $\rho = 766$ AU. At wider separations, the completeness limit of the Gaia catalog ($G \sim 20.5$ mag at moderate galactic latitudes; \citealt{GaiaDr2}) corresponds to an absence of any companions down to a limit of $M \sim 0.015 M_{\odot}$.

\subsection{Spectroscopy}
\targ~was observed multiple times with a range of spectrographs, starting in 2004, and mostly during 2019. In order to derive the stellar properties, we obtained low-resolution spectra on with the SOAR/Goodman spectrograph. To construct the radial velocity curves needed to confirm and further characterize the system, we obtained high resolution multi-epoch spectroscopy with SALT/HRS, the SMARTS/CHIRON, LCO/NRES, and archival data from La Silla 2.2\,m/FEROS spectrographs. Below we describe these observations, and Table \ref{rvtab} lists the instruments used, observation dates, and resulting radial velocity measurements. 

\subsubsection{SOAR/Goodman}

\targ~was observed with the  Goodman High-Throughput Spectrograph \citep{Goodman} on the Southern Astrophysical Research (SOAR) 4.1 m telescope located at Cerro Pachón, Chile. On 2019 August 3 (UT) and under clear (photometric) conditions, we obtained 5 spectra of \targ, each with an exposure time of 15s. We took all exposures using the red camera, the 400 l/mm grating in the M2 setup, and the 0.46\arcsec\ slit rotated to the parallactic angle, which yielded a resolution of $R\simeq$1850 spanning 5000--9000\AA. 

Using custom scripts, we performed bias subtraction, flat fielding, optimal extraction of the target spectrum, and mapping pixels to wavelengths using a 4th-order polynomial derived from the HgArNe lamp spectra. We then stacked the five extracted spectra using the robust weighted mean (for outlier removal). The stacked spectrum had a signal-to-noise ratio $>100$ over the full wavelength range (excluding areas of strong telluric contamination). While we took no spectrophotometric standards during the night, we instead corrected instrument throughput with wavelength using standards from an earlier night.

\subsubsection{SALT/HRS}
\targ~was observed with the Southern African Large Telescope High Resolution Spectrograph (SALT:HRS; \citealt{salt_2006}, \citealt{SALTHRS_2014}) on three nights (2019 July 25 and 28, August 4). On each visit, \targ~ was observed with 350\,$\mu$m fibres in high resolution mode for 320\,s. The HRS data were reduced using the MIDAS pipeline \citep{midas_hrs}, which performs flat fielding, bias subtraction, and wavelength calibration with lamp exposures. The resulting spectral resolution is $R\sim$ 46,000.

\subsubsection{LCOGT/NRES}
\targ~ was observed on six occasions with the Los Cumbres Observatory (LCOGT) telescopes Network of Robotic Echelle Spectrographs (NRES; \citealt{NRES2018}). These are a network of R=53,000 fibre-fed spectrographs on 1\,m telescopes operating in the optical (380-860\,nm). These observations spanned 2019 July 16 to 2019 July 31. Each visit consisted of a 2400\,s exposure, which provided a spectrum with signal to noise of 30-50. The raw  data were automatically reduced with the NRES pipeline. This included bias subtraction, performing order trace and optimal extraction of the spectra, cosmic-ray rejection, and wavelength calibration. 

\subsubsection{SMARTS/CHIRON}
\targ~ was observed on eleven nights spanning 2019 July 30 to 2019 August 8. We obtained 11 epochs of spectroscopic observations with the CHIRON spectrograph on the 1.5\,m SMARTS telescope spanning 2019 July 30 to 2019 August 8. CHIRON is a high resolution echelle spectrograph fed by an image slicer and a fiber bundle, located at Cerro Tololo Inter-American Observatory (CTIO), Chile. The spectra have a resolution of $R=80000$ with a wavelength coverage of $4100-8700$\,\AA{} \citep{2013PASP..125.1336T}. The wavelength solution is provided by bracketing Thorium-Argon cathode-ray lamp observations. The spectra are extracted with the official CHIRON pipeline. The velocities are derived by cross correlations against synthetic templates from the ATLAS9 atmospheric library \citep{Castelli:2004}. 

\subsubsection{La Silla 2.2\,m/FEROS}
\targ~ was observed on 2004 March 11 with ESO's Fibre-fed, Extended Range, Echelle Spectrograph (FEROS) on the La Silla 2.2\,m telescope \citep{kaufer_feros}. These observations were taken as part of Program 072.D-0021(B) (PI Nitschelm) in a search for spectroscopic binaries in Sco-Cen. FEROS provides spectra of R$\sim$48,000 covering 350-920\,nm, whic are reduced with the MIDAS pipeline. This included flat fielding, bad pixel masking, sky-subtraction and wavelength calibration.

\begin{deluxetable}{lccc}
\centering
\tabletypesize{\scriptsize}
\tablewidth{0pt}
\tablecaption{Radial velocity and projected rotational velocity measurements of  host star \targ. \label{rvtab}}
\tablehead{\colhead{Instrument} & \colhead{BJD} & \colhead{RV}  \\
\colhead{} & \colhead{} & \colhead{km/s}}
\startdata
LCOGT:NRES	&58680.52486	 &8.3$\pm$0.5	 \\	
LCOGT:NRES	&58682.53411	 &7.4$\pm$0.4	 \\	
LCOGT:NRES 	&58684.47585	 &8.4$\pm$0.9	 \\	
LCOGT:NRES	&58685.47600	 &7.7$\pm$0.6	 \\	
LCOGT:NRES	&58692.47785 	 &7.9$\pm$1.1	 \\	
LCOGT:NRES	&58695.56019 	 &8.1$\pm$0.8	 \\	
SALT:HRS	&58690.30040 	 &7.15$\pm$0.19. \\
SALT:HRS    &58693.29021 	 &7.40$\pm$0.18. \\
SALT:HRS	&58700.28479 	 &7.42$\pm$0.07. \\
SMARTS:CHIRON &58694.61145   & 7.26$\pm$0.17 \\
SMARTS:CHIRON &58695.59035   & 7.64$\pm$0.26 \\
SMARTS:CHIRON &58698.55321   & 7.72$\pm$0.18 \\
SMARTS:CHIRON &58700.53088   & 7.93$\pm$0.22 \\
SMARTS:CHIRON &58701.47900   & 7.59$\pm$0.19 \\
SMARTS:CHIRON &58702.54990   & 7.40$\pm$0.14 \\
SMARTS:CHIRON &58704.48051   & 7.95$\pm$0.16 \\
SMARTS:CHIRON &58705.53525   & 7.60$\pm$0.16 \\
SMARTS:CHIRON &58706.50439   & 7.53$\pm$0.2  \\
SMARTS:CHIRON &58707.46434   & 7.60$\pm$0.17 \\
SMARTS:CHIRON &58708.48880   & 7.19$\pm$0.27 \\
ESO\,2.2m:FEROS & 53075.78544& 8.58$\pm$0.07 \\
\multicolumn{2}{r}{Error Weighted Mean RV}&7.78$\pm$0.54\\
\multicolumn{2}{r}{SALT~~$v\sin{i}$}     & 54.9$\pm$0.1\\
\multicolumn{2}{r}{LCOGT:NRES~~$v\sin{i}$} & 54.9$\pm$0.4\\
\multicolumn{2}{r}{SMARTS:CHIRON~~$v\sin{i}$} & 51.9$\pm$0.3\\
\multicolumn{2}{r}{ESO\,2.2m:FEROS~~$v\sin{i}$} & 54.2$\pm$0.02\\
\multicolumn{2}{r}{Error Weighted Mean~~$v\sin{i}$}&54.2$\pm$0.7\\
\enddata
\end{deluxetable}

\subsection{Literature Photometry and Astrometry}\label{sec:litstuff}
To characterize the properties of the host star, we drew photometry and astrometry from the wider literature for \targ. We took proper motions, parallax, and optical $G$, $RP$, and $BP$ photometry from the Gaia second data release (DR2; \citealt{GaiaDr2}). We also drew $V_T$ and $B_T$ photometry from Tycho 2 \citep{perrymantycho}, $B$ and $g$' photometry from the American Association of Variable Stars Observers All-Sky Survey (APASS; \citealt{apass}), near infrared $J$, $H$, and $K_s$ photometry from the Two Micron All Sky Survey (2MASS; \citealt{2mass}), and mid infrared $W_{1}-W_{4}$ photometry from the Wide-Field Infrared Survey Explorer (AllWISE, \citealt{wise10,wise_mainzer}).

\section{Measurements}\label{sec:measure}
\subsection{Stellar Parameters}\label{sec:stelpars}
\emph{Luminosity, Radius, and Effective Temperature:}

We fit the  spectral-energy-distribution of \targ~using the available photometry, the Goodman optical spectrum, and spectral templates of nearby (unreddened) young stars. More details of our method can be found in \citet{Mann2015b}, with additional details on how we fit for reddening in \citet{zeit3}, both of which we summarize here. We compared the photometry described in Section \ref{sec:litstuff} to synthetic magnitudes derived from the combination of our SOAR spectrum, the template spectra, and Phoenix BT-SETTL models \citep{Allard2011} to cover gaps in the spectra (e.g., beyond 2.5\um), using appropriate filter profiles and zero points \citep[e.g.,][]{Cohen2003, BessellMurphy2012, Mann2015a, dr2_filter}.  Errors in the photometry account for measurement errors, uncertainty in the zero-points and filter profiles (where available), and an estimate of stellar variability (0.02\,mag in the optical). Photometry was weighted by these uncertainties in the comparison. We included reddening as an additional free parameter (assuming $A(V)/E(B-V)=3.1$), which is added to the template and model spectra to better match the observed quantities. Tests against external spectra suggest our flux calibration is only good to $\simeq$10\%, so we included two additional free parameters to model errors in the spectral shape. 

\begin{figure}
    \centering
    \includegraphics[width=0.49\textwidth]{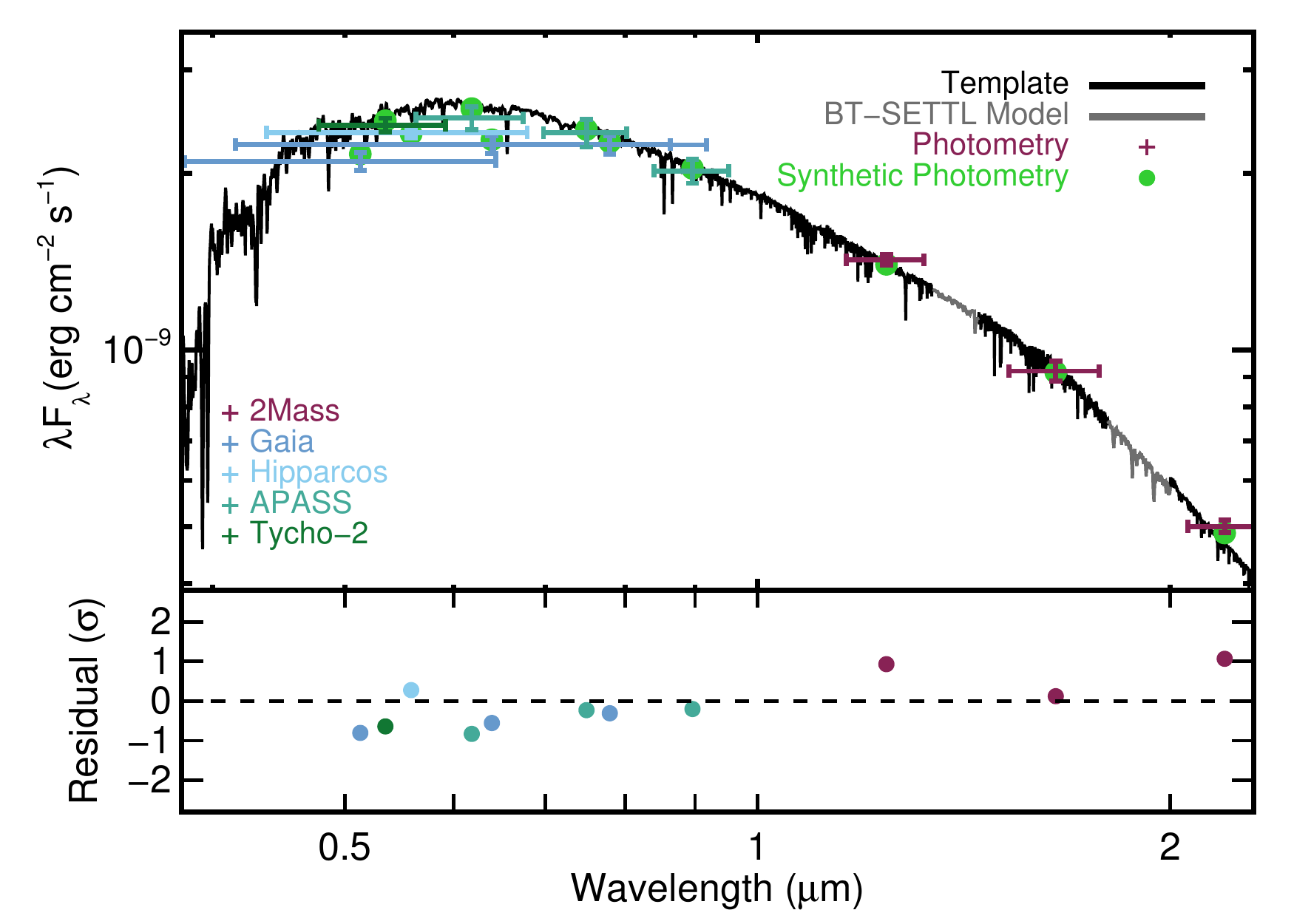}
    \caption{Best-fit spectral template and Goodman spectrum (black) compared to the photometry of \targ. Grey regions are BT-SETTL models, used to fill in gaps or regions of high telluric contamination. Literature photometry is colored by the source, with horizontal errors corresponding to the filter width and vertical errors the measurement errors. Corresponding synthetic photometry is shown as green circles. The bottom panel shows the residuals in terms of standard deviations from the fit, matching the color of the photometry source.}
    \label{fig:sed}
\end{figure}

 In addition to E(B-V), our $\chi^2$ comparison provided and estimate of $T_{eff}$ from the the best-fit atmosphere model, $L_*$ from the integral of the calibrated spectrum and Gaia DR2 distance, and $R_*$ from the Stefan-Boltzmann relation. For \targ, our final values were $T_{eff} = 5650 \pm 75$\,K, $L_* = 1.75 \pm 0.09 L_\odot$, and $R_* = 1.392 \pm 0.055 R_\odot$, with a low extinction of ($E(B-V)=0.02^{+0.03}_{-0.02}$). The final calibrated and combined spectrum is shown in Figure~\ref{fig:sed}. 

\emph{Age and Stellar Mass:} We fit the effective temperature and luminosity of \targ~ measured in Section \ref{sec:stelpars} to the PARSEC 1.2s isochrones of \citet{bressan12} and the BHAC15 isochrones of \citet{bhac15} to determine an age and mass estimate for the host star. We did not adopt an age {\it a priori}, as the age of stars in different regions of the Sco-Cen association can range from 5-20\,Myr \citep{pecaut16}. We fit a model with age and mass as parameters. We interpolated the isochrone grid and explored the posterior using the MCMC framework {\tt emcee} \citep{emcee}. In our fit, we assumed Solar metallicity for \targ, which is consistent with estimates for the young regions around Sco-Cen \citep{james06}.

We find a best fit age of  $\tau$=17.0$^{+2.7}_{-1.5}$\,Myr for the BHAC15 models and $\tau$=16.6$\pm$2.2\,Myr for the PARSEC models. Both models produced similar masses of M=1.2$\pm$0.05\,$M_\odot$. In combination with our stellar radius measurement, this gives a stellar density $\rho_\star/\rho_\odot = 0.46\pm0.06$ which we adopt as a prior for the transit analysis described below. 

\subsection{Radial Velocities}\label{sec:rvs}
To determine radial velocities (RVs), we compute spectral-line broadening functions (BFs; \citealt{Rucinski1992,Tofflemireetal2019}). BFs were computed using a 6000\,K synthetic template \citep{husser13}, which was found to provide the best order-to-order RV stability from a grid of templates sampled in steps of 100 K. Adjacent templates within 300 K provided consistent RV measurements. In each epoch we find the BF is rotationally broadened, but single-peaked, indicating the contribution from a single star.

To measure RVs the BF is computed for each echelle order and combined, weighted by its S/N, and fit with a rotationally-broadened stellar line profile. Uncertainties on these measurements are calculated as the standard deviations of the profile fits from BFs combined from three independent subsets of the echelle orders. Some epochs consisted of three consecutive integrations, for which the quoted RV and uncertainty represent the error weighted mean and standard error of the three individual spectra. 

Table \ref{rvtab} presents our RV measurements. We found a mean radial velocity of 7.51$\pm$0.23\,km/s, which is consistent with the value of 8.1$\pm$1.0\,km/s found by \citet{TorresSearch2006}, but inconsistent with the more recent value of -2.3$\pm$0.4\,km/s reported by \citet{chen11}. The discrepancy with the latter may be the result of the source's large rotational broadening (see Section \ref{vsini}), which results in Fourier cross-correlation functions that are not well fit by Gaussian profiles. 

\subsection{Projected Rotational Velocity}
\label{vsini}
As a free parameter of the stellar line profile fit above, we also measure the projected rotational velocity $v \sin (i)$. From the mean and standard error for each data set, we measure a value of 54.9$\pm$0.1\,km/s for the SALT:HRS spectra, and 54.9$\pm$0.4\,km/s from LCOGT:NRES. These fits do not include broadening contributions from micro-turbulence, but that is likely to have a small effect given the large $v \sin (i)$ measured. From the SMARTS:CHIRON observations we measure a value of 51.9$\pm$0.3\,km/s, and with FEROS we measure $v \sin (i)$ of 52.20$\pm$0.02\,km/s. These values are all consistent with the values reported by \citet[][51\,km/s]{TorresSearch2006} and  \citet[][57$\pm$3\,km/s]{chen11}, and the variance between measurements likely reflects the treatment of broadening sources by the different analyses.

\subsection{Stellar Rotation}
The \tess~lightcurve of \targ~ shows constant modulation of 1-2\% peak-to-peak amplitude on timescales shorter than 2\,days. To determine the rotation period, we took the \tess~lightcurve, masked out the transits of the candidate planets from the data, and computed a Lomb-Scargle periodogram spanning periods of 1–15 days. There are peaks in the periodogram at $\sim$0.71\,days and $\sim$1.4\,days, the latter of which is consistent with period measurements from ground-based observations from \citet{mellon17}. The lightcurve morpohology in the second half of the \tess{} dataset clearly indicates that the longer period is associated with the rotation of the star.


To determine the credible interval on the stellar rotation period, we fit the \tess{} lightcurve with a Gaussian process using the \texttt{celerite} package \citep{Foreman-MackeyFast2017}. We used a kernel composed of two stochastically-driven, damped harmonic oscillators, with two Fourier modes; one at the rotation period of the star, and one at half the rotation period. This was done as part of our transit fitting methodology described in Section \ref{sec:transfit}. We measure the rotation period to be 1.418$\pm$0.016\,days. 

Given the measurements of the stellar rotation period, projected rotation velocity, and the stellar radius, we can estimate both the equatorial velocity and hence the stellar inclination relative to our line of sight and the orbit of the transiting planet candidate. We compute an equatorial velocity for \targ~ of  49.2$\pm$2.2\,km/s and use the Bayesian methodology of \citet{morton14} to compute the likelihood of the stellar inclination. We find a loose constraint of $i>66^\circ$ at 99\% confidence, noting that we adopt $i<90^\circ$ here, though we cannot distinguish between $i<90^\circ$ and $i>90^\circ$.

\section{Membership in the Sco-Cen Association}\label{sec:membership}
\targ~was first identified as a member of the Upper-Centaurus-Lupus (UCL) subgroup of the Sco-Cen association on the basis of proper-motion co-movement with other association members by \citet{zeeuw99}. \targ~also features multiple spectral indicators of youth that support membership in Sco-Cen. In particular, \targ~appears lithium rich for an early G dwarf, and exhibits X-ray and  H-$\alpha$ emission typical of other young stars in Sco-Cen \citep{mamajek02,chen11}. 

We reassess the kinematic association of \targ~ with Sco-Cen on the basis of the new Gaia DR2 astrometry and photometry, and our radial velocity monitoring. The three dimensional space motion of \targ~is $(U,V,W)=(7.12,-21.53,-5.67)\pm(0.19,0.22,0.13)$\,km/s. This is highly consistent with the velocity of UCL \citep{wright18}. The distance of $\sim$127\,pc also places \targ~ within $1\sigma$ of the Sco-Cen median distance ($d=110\pm20$\,pc) in that region of the association. We compute a membership probability for \targ{} in Sco-Cen of 96\% following the Bayesian prescription of \citet{myfirstpaper} and using the Gaia DR2 astrometry and the radial velocity measured in Section \ref{sec:rvs}. \targ\ is therefore a high confidence kinematic member of the association.

\begin{figure}
    \centering
    \includegraphics[width=0.49\textwidth]{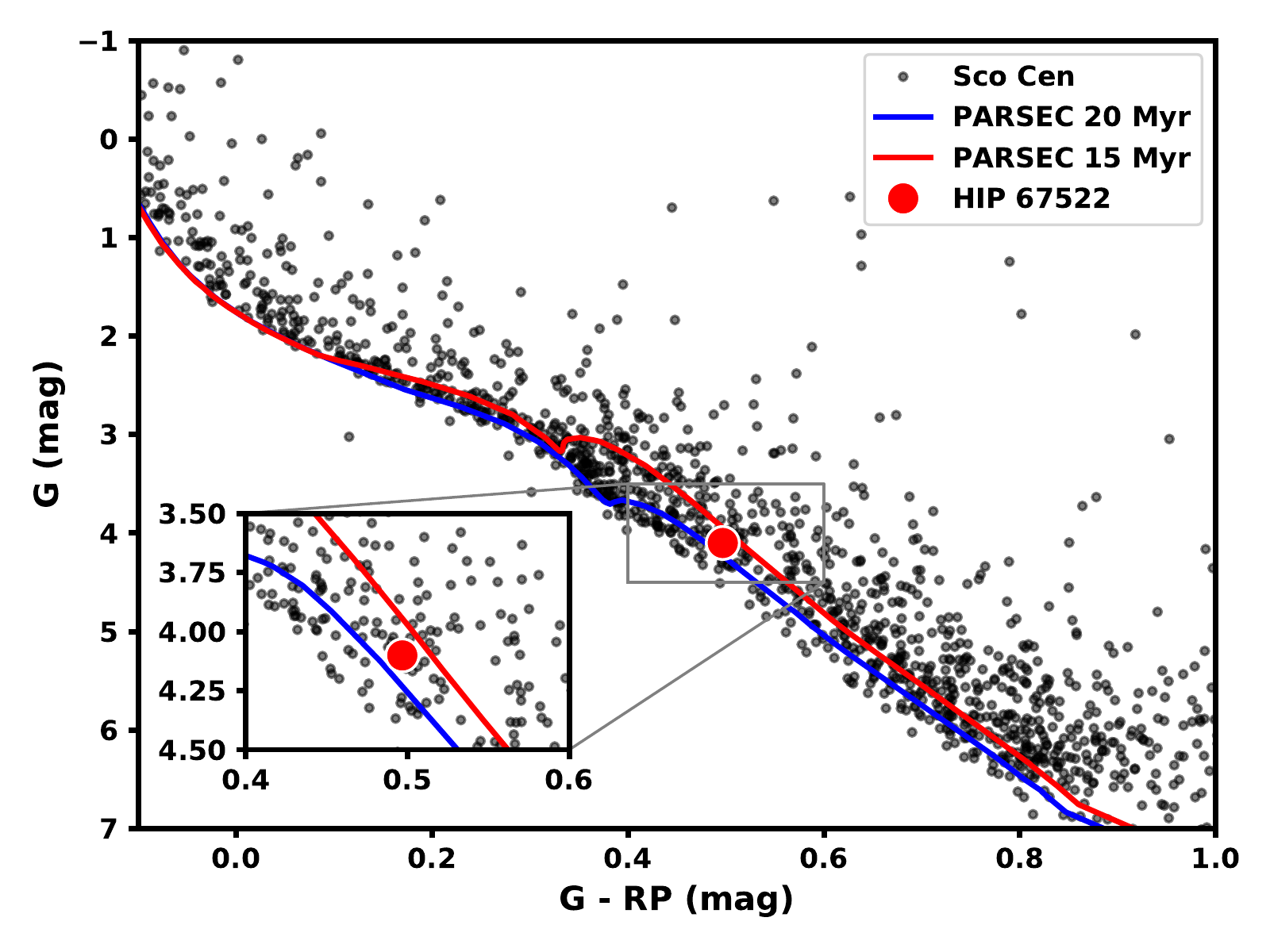}
    \caption{The color-magnitude diagram for high-probability kinematic members of the Sco-Cen association, relative to \targ~(red circle), produced using Gaia DR2 photometry and parallaxes \citep{GaiaDr2}. The red and blue lines are the PARSEC 1.2s \citep{bressan12} 15 and 20\,Myr isochrones respectively. \targ~sits between the two isochrones, as expected for a member of UCL.}
    \label{fig:CMDplot}
\end{figure}

The Color-Magnitude Diagram position of \targ~ places it in the association sequence. Figure \ref{fig:CMDplot} shows the Gaia ($R_P-G,G$) CMD for high probability Sco-Cen members, as well as isochrones from the PARSEC 1.2s models for ages of 15 and 20\,Myr \citep{bressan12}. \targ~ falls between the two models isochrones, as is expected for UCL stars with ages ranging from 15-20\,Myr.

\section{Transit Fitting}\label{sec:tfits}

\subsection{HIP~67522\, b}{\label{sec:transfit}}

\begin{figure*}
    \centering
    \includegraphics[width=0.99\textwidth]{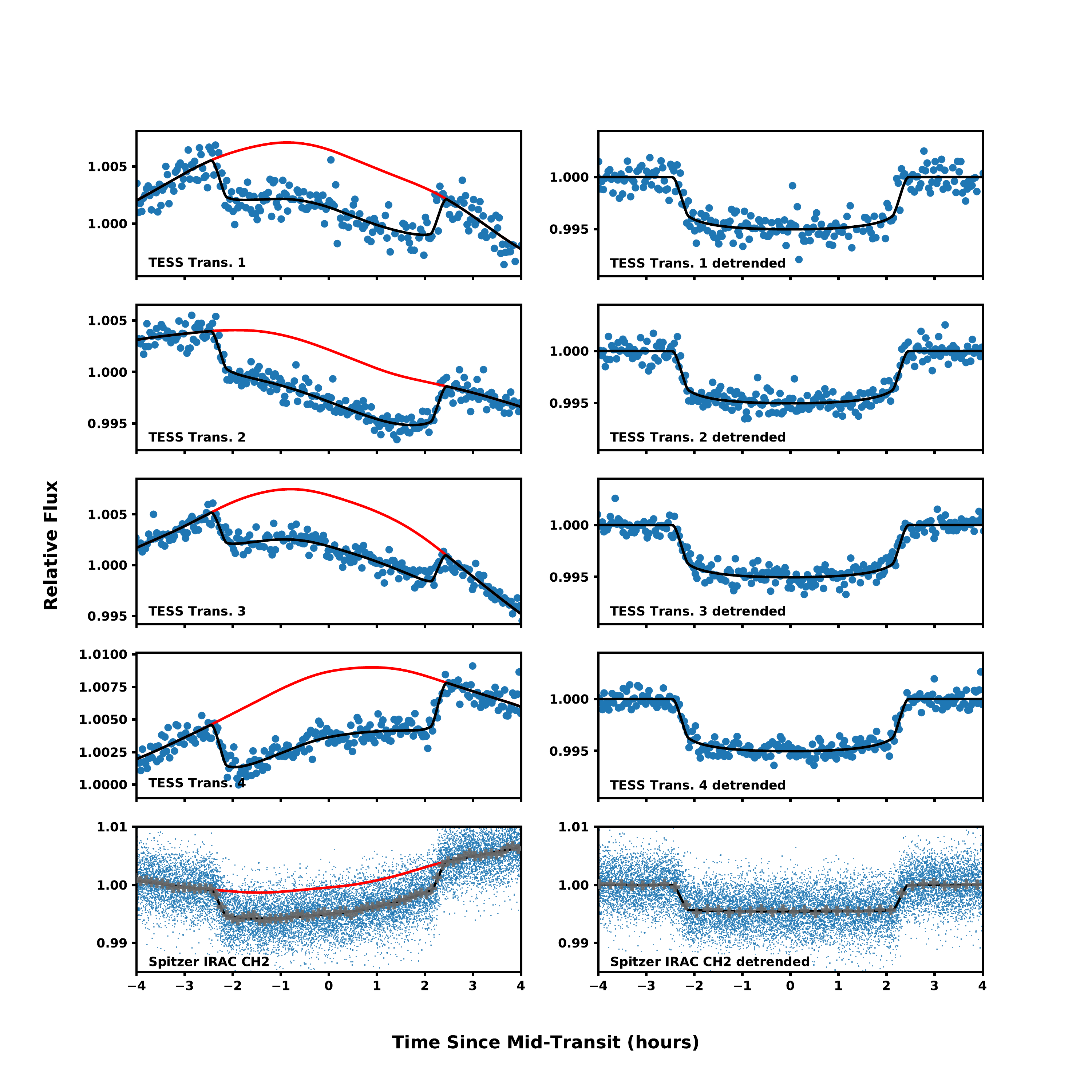}
    \caption{The five transits of \targ\,b, four from \tess{} and one from \Spitzer{} IRAC channel 2 (corrected using BLISS), with and without Gaussian process detrending. (\emph{Left Column:}) The transits without variability removed using the GP model. Blue points are the photometry, the black line is the full transit and GP model, and the red line is the GP model only. (\emph{Right Column:}) The same data with the GP model component removed. The black line now shows the best-fit transit model. In the two plots containing \spitzer{} data, the grey pluses are the data binned into 70 bins.}
    \label{fig:transitsfit}
\end{figure*}

\begin{figure*}
    \centering
    \includegraphics[width=\textwidth]{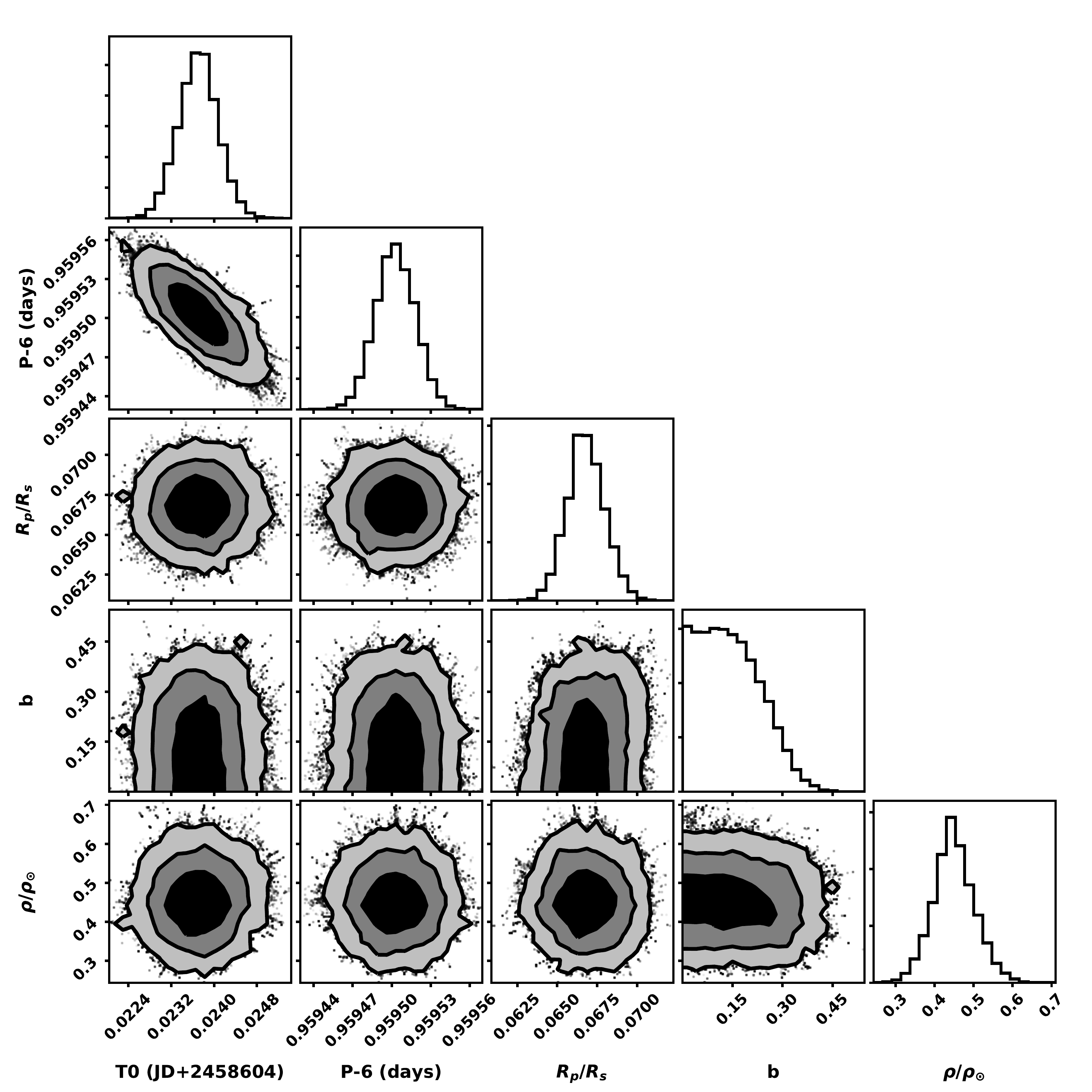}
    \caption{Selected transit fit posteriors for HIP 67522\,b.}
    \label{fig:transcorner}
\end{figure*}

\begin{figure}
    \centering
    \includegraphics[width=0.5\textwidth]{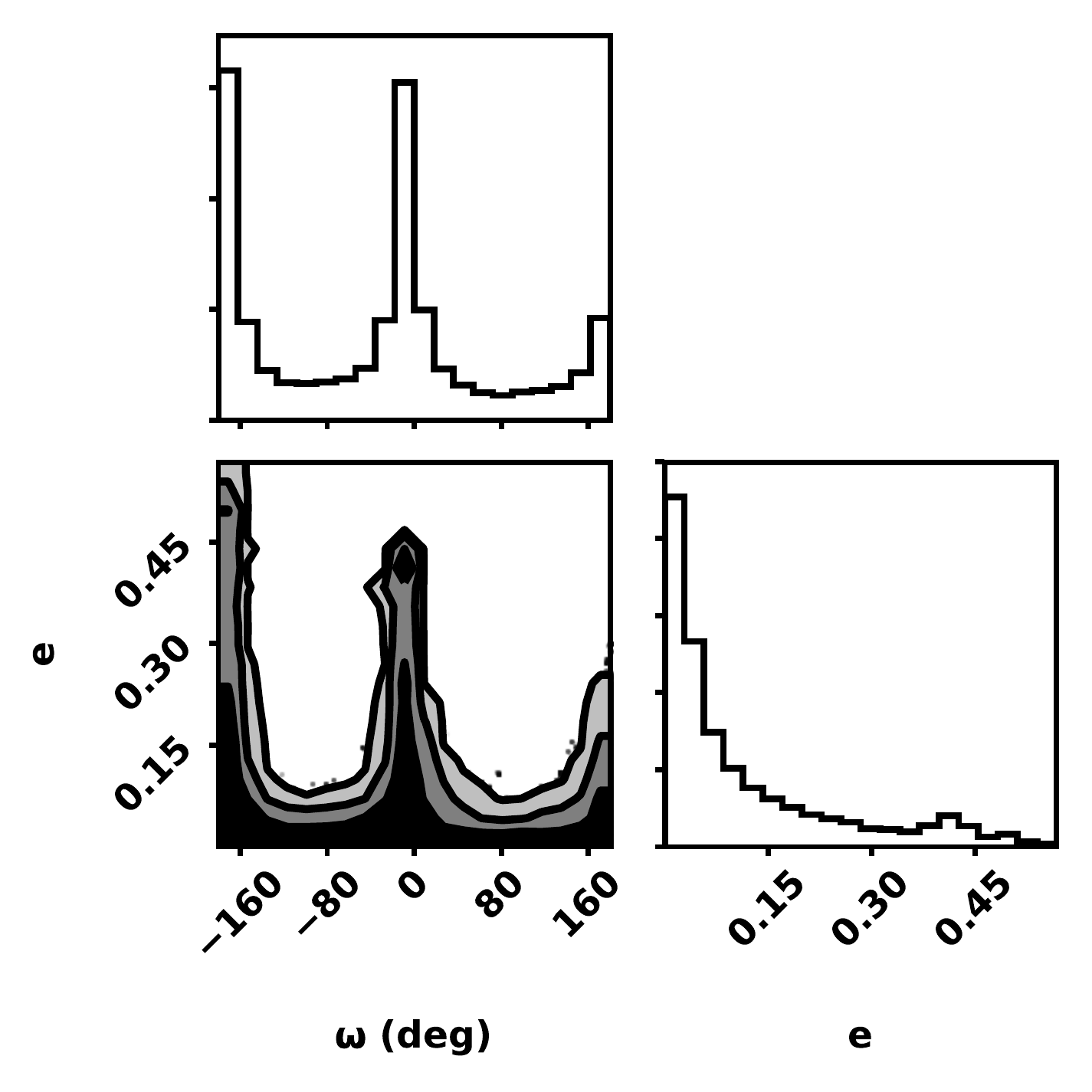}
    \caption{Transit fit posteriors for eccentricity and argument of periastron ($\omega$).}
    \label{fig:eomega_corner}
\end{figure}

We fit a combined transit and variability model to the \tess{} and \spitzer{} photometry using the \texttt{misttborn} \citep{JohnsonK22602018} transit fitting code\footnote{\url{https://github.com/captain-exoplanet/misttborn}}, with the single transit candidate and any flares identified by inspection masked from the lightcurve. The transit model was generated with the BAsic Transit Model cAlculatioN (BATMAN; \citealt{batman}) model to generate model lightcurves, and the stellar rotational variability was modelled with a Gaussian Process (GP) using the {\tt celerite} package \citep{Foreman-MackeyFast2017}. We chose a GP kernel that reasonably matched the rotational periodogram power spectrum, which is dominated by the rotation period and the first harmonic. Specifically, the kernel consisted of two stochastically-driven, damped harmonic oscillators, one at the model rotation period, and the second at the first harmonic (or half the rotation period) and a jitter term to account for the possibility of additional stochastic errors. The amplitude and oscillator quality factor of each component oscillator was allowed to vary, though the primary component was forced to be less damped than the first harmonic in all cases. We then used the \texttt{emcee} \citep{emcee} affine-invariant MCMC package to explore the parameter space. The transit model described above consists of the following parameters: time of periastron ($T_0$), orbital period of the planet ($P$), planet-to-star radius ratio ($R_p/R_\star$), impact parameter ($b$), two eccentricity parameters $\sqrt{e}\sin{\omega}$ and $\sqrt{e}\cos{\omega}$, where $e$ is the orbital eccentricity and $\omega$ is the argument of periastron, the stellar density ($\rho_\star/\rho_\odot$), the linear and quadratic limb-darkening coefficients for \tess{} ($q_{1,1}$,$q_{2,1}$) as per the triangular sampling prescription of \citet{Kipping2013}, and the corresponding limb-darkening coefficients for \spitzer{} ($q_{1,2}$,$q_{2,2}$). The GP parameters, all explored in log-space, were the variability amplitude of the fundamental oscillator ($\log A_1$), the quality factor of the half-period oscillator ($\log Q_{2}$), the difference in quality factors for the two oscillators ($\Delta Q = \log{Q_1}-\log{Q_2}$), the rotation period ($\log P_R$), and a mixture term describing the amplitude ratio of the two oscillators ($m$, where $A_1/A_2 = 1+e^{-m}$). The mixture term, $m$, was formulated in this way to facilitate sampling without the requirement for bounds while ensuring the oscillator at the full rotation period had the largest amplitude. Finally, the stochastic jitter term had a single parameter ($\log{\sigma_j^2}$) which represented the variance of the jitter model.

We applied Gaussian priors on the limb-darkening coefficients for both \tess{} and \spitzer{} based on the values in \citet{ClaretAstronomy2011} and \citet{2017A&A...600A..30C}. We also used Gaussian priors for the stellar density, taken from our derived stellar parameters in Table \ref{proptab}. All other parameters were sampled uniformly, with physically motivated boundaries, namely  $\sqrt{e}\sin{\omega}$ and $\sqrt{e}\cos{\omega}$ bound by $(-1,1)$, and $|b|<1+R_p/R_s$. The GP parameters were allowed to vary without bounds on the real line, except for $\Delta Q$ which was set to be larger than zero.  We used 50 walkers in the MCMC, running 240,000 steps each, with a burn-in of 120,000 steps.  These chain lengths correspond to 100, and 60 times the integrated autocorreltaion times respectively s described in \citet{goodman_weare2010}.  The resulting fit is shown in Figure \ref{fig:LCplot} with the individual transits shown in figure \ref{fig:transitsfit}, and the best fitting model and derived parameters, along with 68\% credible intervals are listed in Table \ref{tab:transfit}. Figures \ref{fig:transcorner} and \ref{fig:eomega_corner} show the posteriors for the transit fit parameters.

\subsubsection{False Positive Probability}\label{sec:falsep}

Given the $\sim$2 arcsecond resolution of the \spitzer{} imaging, it is possible that the transit signals of \targ\,b are caused by some astrophysical system other than a transiting exoplanet. We directly address some of these false positive scenarios for \targ\,b as follows:

\begin{enumerate}
    \item \emph{The transits are caused by instrumental artifacts or
    residuals from stellar variability}: Though the four transits of HIP~67522\,b in the \tess{} dataset have amplitudes
    much smaller than the amplitude of starspot variability, we confirm the transits with \spitzer{}, conclusively.

    \item \emph{\targ~ is an eclipsing binary or brown dwarf}: We generated 1,000,000 binary companions with a uniform distribution in all orbital parameters except a Gaussian distribution in period following \citet{Raghavan2010} and the inclination restricted to values where the companion would eclipse the primary ($i\simeq90^\circ$). We then compared the predicted radial velocities to those we measured from our high-resolution spectroscopy (Section~\ref{sec:rvs}), rejecting systems that disagree with the data at $>3\sigma$. For this comparison, we added in a conservative 150\,\ms\ jitter (in quadrature) to all points to account for the star's activity levels. All companions with non-planetary masses and periods below 80\,d can be ruled out. We set a $2\sigma$ limit on the mass of the planet of $M_p<5M_J$.
    
    \item \emph{The transits are blended from a background, unassociated eclipsing binary system or transiting exoplanet:} We take the same approach outlined in \citet{vanderburg2019}. In short, if the transits are a blend from the background system, the true radius ratio is constrained by the ratio of the ingress time (T$_{12}$) to the duration from first to third contact (T$_{13}$), and we can estimate the largest possible magnitude difference between \targ~ and a background object that could produce the transits as $\Delta m \leq 2.5\log_{10}(t^2_{12}/t^2_{13}/\delta)$, where $\delta$ is the observed transit depth. We repeat the transit fitting described in Section \ref{sec:transfit} with the prior on stellar density and limb-darkening removed to explore the allowed posterior for the duration terms. This yields $\Delta m < 2.6$\,magnitudes at 99.9\% confidence. Interpolating on the \citet{bhac15} 2\,Gyr isochrone, this is equivalent to a companion of mass $>$0.74\,M$_\odot$. The imaging and astrometric limits to companions presented in Table \ref{tab:contrastlims} rule out such objects down to $\sim$60\,mas.

    \item \emph{\targ~ is a hierarchical triple}: We generate an additional 1,000,000 companions as above, but now assign orbital periods based on a log-normal distribution following \citet{Raghavan2010}. We compare each binary to the AO, radial velocity, and Gaia derived constraints on companions, adding jitter to the velocity data as in the first scenario. We exclude companions too faint to reproduce the observed transit depth (following the method above) and those that would be resolved in the \spitzer{} imaging. Because bound companions would be younger and brighter than a typical field star, the constraints are stronger than the background eclipsing binary case; less than 1\% of the generated companions survive. Survivors that could have reproduced the transit depth are generally companions that happen to be aligned with the host star in the plane of the sky. Considering the underlying rate of hierarchical triples is $\simeq$16\% \citet{Raghavan2010}, this scenario is ruled out statistically. We do not include constraints from chromaticity of the transit depths (\tess{} versus \spitzer), which would strengthen the constraints \citep{Desert2015}. 

\end{enumerate}

We quantify the likelihood of one of the above scenarios causing the transits of \targ\,b using the open source {\tt vespa} software package \citep{Morton2015}. {\tt Vespa} computes the false positive probability (FPP) according to \citet{Morton2012} and \citet{Morton2016}. This is done with a Bayesian model comparison between various scenarios that may cause transit-like signals (such as an exoplanet, background eclipsing binaries, eclipsing binary on the primary, and eclipsing binary on a companion) using the transit photometry, observational constraints on companions, and the host star parameters and photometry. We ran {\tt vespa} with the stellar parameters of \targ~in Table \ref{proptab}, the \tess{} lightcurve with the GP model described in Section \ref{sec:tfits} subtracted, and the constraints to companions from the speckle imaging and Gaia astrometry from Table \ref{tab:contrastlims}. Based on these inputs, {\tt vespa} calculated a false positive probability $FPP<1\times10^{-6}$ for  \targ\,b. We hence consider \targ\,b to be a validated exoplanet.

\subsection{Single Transit Parameters}\label{sec:singletrans}
We model the single transit with the same methodology as was used for the periodic signal in Section \ref{sec:transfit}, with modification to allow for an unconstrained orbital period. We use the same GP to model the out of transit rotation variability, but constrain the parameters with Gaussian priors with standard deviation equal to the 68\% credible intervals from the periodic fit shown in Table \ref{tab:transfit}. Similarly, we constrain the limb darkening parameters in the \tess{} bandpass, $q_1$ and $q_2$, and the stellar density $\rho_\star/\rho\odot$ based on the previous fit. We allowed the orbital parameters to vary as in Section \ref{sec:transfit}, and applied an unbiased prior on eccentricity taken from \citet{kipping14}.

 \begin{figure}
     \centering
     \includegraphics[width=0.48\textwidth]{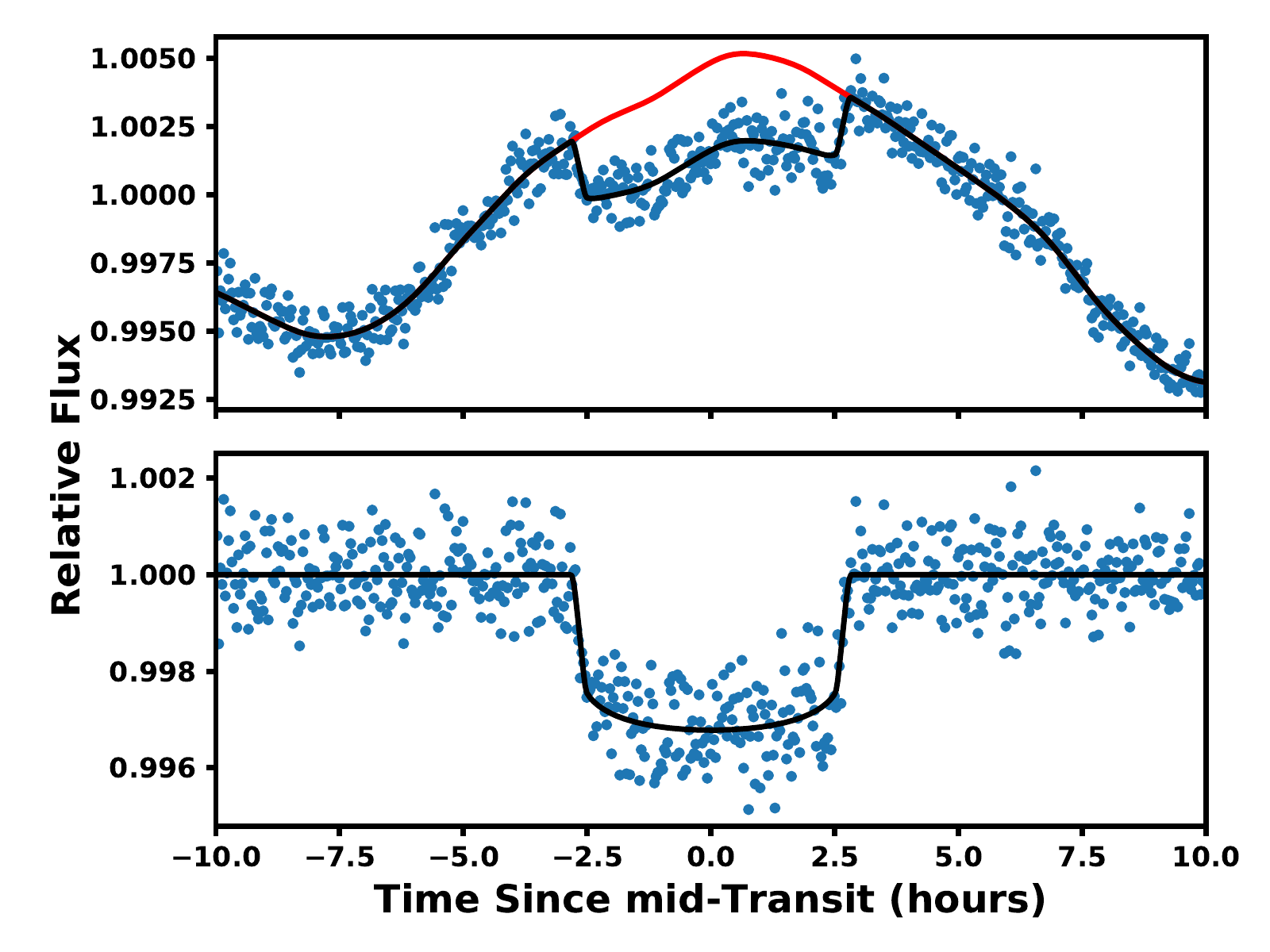}
     \caption{Transit of the singly transiting object and best-fit model. The upper panel shows the PDC lightcurve (blue points) with the GP (red), and GP with transit model (black), while the lower panel shows the same with the GP model removed.}
     \label{fig:singlefit}
 \end{figure}

Beyond the transit shape, we also have some information regarding the orbital period of the candidate exoplanet based on the lack of an additional transit in the \tess{} Sector 11 dataset. Similarly to \citet{VanderburgZodiacal2018}, we apply this as a prior in the MCMC to the estimated periods based on the time baseline of the \tess~ lightcurve for \targ;
\begin{equation}
    P(P_{single}) \propto (\tau + t_{14})/P_{single},
\end{equation}
 where $\tau$ is the time baseline of the \tess~ observation in sector 11 (23\,days). We explored our model posterior using an MCMC framework and the {\tt emcee} package \citep{emcee}. We used 40 walkers and iterated for 200,000 steps with a 100,000 step burn-in. Figure \ref{fig:singlefit} shows the single transit and best-fit model, and Table \ref{tab:transfitc} lists the best-fit model parameters.

\section{Discussion}\label{sec:discussion}

We have reported the detection and characterization of a 17\,Myr old transiting giant planet, HIP 67522\,b in the Sco-Cen association, with an orbital period of 6.9596\,days and a candidate second planet in a much longer period orbit. HIP 67522\,b is currently the youngest confirmed transiting, hot, Jupiter-sized planet, pending the confirmation of the possible hot-Jupiter PTFO~8-8695\,b (CVSO-30\,b) \citep{vaneyken,yuptfo,johnskrull_ptfo}. \targ\,b is one of only a small handful of exoplanets younger than 20\,Myr (K2-33\,b \citealt{zeit3}; V1298 Tau bcde; \citealt{trevorv1298}) and as such may be significant as a benchmark young system in future models of exoplanet formation and migration.

\targ\,b has a radius measured from the transit data from \tess{} and \spitzer{} of R$_P$=10.02$^{+0.54}_{-0.53}$\,R$_\oplus$. The exoplanet mass-radius relationship of \citet{Chen17_forecaster} and the constraints to planet masses given in Section \ref{sec:falsep}, predict a mass in the range of $0.18-7.2$\,M$_J$. Our radial velocity monitoring in Section \ref{sec:rvs} indicates a mass M$_P<5$\,M$_J$, ruling out the highest part of this range. Given the youth of the system, a mass radius relation drawn from mature exoplanet systems with ages $<$1\,Gyr may not directly apply. In particular, the radius of \targ\,b may be inflated for its true mass, as has been suggested for members of open clusters with ages $<$1\,Gyr (e.g., \citealt{zeit4,RizzutoZodiacal2018})

The transits of \targ\,b are unlikely to be caused by a non-planetary phenomenon; we calculate a false positive probability using {\tt vespa} of 1$\times$10$^{-6}$. From a combination of speckle imaging, and analysis of the astrometric data from the Gaia DR2 catalog, we find no evidence for stellar or brown dwarf companions down to $\sim$100\,mas. High resolution spectroscopic monitoring of \targ~rules out spectroscopic companions with masses larger than 13\,M$_J$. 

We also detect a single transit of a third body in the \targ~system in the \tess{} lightcurve which we will tentatively refer to as \targ\,c. From fitting transit models to the single transit of \targ\,c, we find a radius of 8.01$^{+0.75}_{-0.71}$\,R$_\oplus$, and an orbital period that must be larger than $\sim$24\,days. The upcoming \tess{} first extended mission \citep{huang2018}, which will revisit much of the Sco-Cen association may provide confirmation of \targ\,c in Sector 38 if the true orbital period is not significantly longer than the typical \tess{} observing sector length. We calculate the probability of \tess{} observing a transit if \targ\,c in Sector 38 based on the period posterior from our transit model in Setion \ref{sec:singletrans} to be $\sim$54\%.

The constraints from the stellar parameters of \targ{}, and the duration of the transits in the \tess{} and \spitzer{} lightcurves imply zero eccentricity for the orbit of \targ\,b (Figure \ref{fig:eomega_corner}). Large eccentricities ($>$0.3) are strongly ruled-out (Figure \ref{fig:eomega_corner}). The stellar rotation axis is also weakly constrained to be aligned to the orbital plane. In combination with the presence of the outer planet \targ\,c, this suggests possible formation or migration mechanisms responsible for the architecture of this system. In particular, the existence of a short-period hot-Jupiter at 17\,Myr itself means that it is possible to produce a short-period gas-giant while still on the early pre-main-sequence. Disk migration can induce migration to short periods on such timescales \citep{lubow_migration}, and disk-planet interactions are not expected to produce eccentricities larger than e$\sim$0.05-0.1 \citep{dunhill2013}. Indeed, disk migration is expected to produce gas giants in short period orbits ($<$20\,days) on timescales shorter than the age of \targ~ with close to zero eccentricities \citep{Cloutier2013}. This scenario is consistent with the transit parameters we have measured for \targ\,b. 

The presence of a second planet on a transiting orbit with period $<$ 100\,days, as is likely for \targ\,c pending confirmation, is somewhat rare for a hot Jupiter system \citep{wright2009,szabo2013} though some examples exits (e.g. \citealt{VanderburgPrecise2017}). This architecture may suggest either ongoing dynamical interactions, or that \targ\,b is in fact less massive than its radius implies. Dynamical interactions between planets, such as planet-planet scattering \citep{ChatterjeeDynamical2008}, and the Kozai-Lidov interaction \citep{fabrycky07} are expected to produce eccentric and often misaligned warm/hot Jupiters. The timescales involved for these interaction are typically 100-1000\,Myr, longer than the $\sim$17\,Myr age of \targ. Given the presence of \targ~c, planet-planet interactions may be important in this system, and orbital evolution of the \targ\,b-c system may still be ongoing.

\subsection{Prospects for follow-up}
The relative brightness ($V = 9.8$ mag) and proximity ($d = 127$ pc) of \targ~make it highly amenable to follow-up, both with ground-based and space-based observatories. Measuring the mass of \targ\,b would provide a rare young-age density measurement, and would allow comparison to older counterparts. Given the mass estimate from \citet{Chen17_forecaster} of $M=0.18-4.6$\,M$_J$, the expected radial velocity amplitude will be in the range of $17-400$\,ms$^{-1}$. At the upper end, this is comparable to the expected RV variability due to rotation of 150\,m s$^{-1}$, calculated using $\Delta RV = \sigma_{phot}v\sin{i}$, where $\sigma_{phot}$ is the photometric variability amplitude in the same band. It will certainly be possible to either detect or rule out the most massive scenarios for \targ\,b with future radial velocity measurements. A dedicated campaign designed to observed and model the radial velocity systematics, perhaps in the NIR where variability is expected to be smaller, may allow a mass measurement even for the smaller side of the possible mass values (e.g., \citealt{k2100_mass}). 

A measurement of the sky-projected angle between the orbital rotation and stellar rotation angular momentum vectors via observation the Rossiter-McLaughlin (RM) effect for \targ\,b is also possible. We estimate the size of the radial velocity amplitude due to the Rossiter-McLoughlin effect to be $\sim$160\,ms$^{-1}$ using the relation $\Delta RV_{RM} = 0.65v\sin{i}(R_p/R_\star)^2\sqrt{1-b^2}$ \citep{gaudi2007}, making this one of the most amenable targets for RM follow-up. Constraining the star-planet obliquity of this young system may shed light on potential migration of dynamical evolution processes responsible for its production (e.g., \citealt{zhoudstuc}, \citealt{montetdstuc}). 

\targ\,b is also a strong candidate for atmospheric characterization with JWST. We compute the transmission spectroscopy metric as described in \citet{kemptontsm} to be $\sim$190, assuming zero albedo and full day-night heat redistribution to derive an equilibrium temperature of $T_{eq}=1174\pm21$\,K, and a mass of 0.25\,M$_J$ as per the prescription of \citet{kemptontsm,Chen17_forecaster}. This can be interpreted as a transmission spectrum measurement with signal to noise of $\sim$190 in a 10 hour window with the NIRISS instrument \citep{Louie_2018}. This places \targ\,b in the top quartile of transiting gas giants for JWST atmospheric characterization. Combined with its young age, this makes \targ~a compelling target for further characterization.

\section{Acknowledgements}
ACR was supported as a 51 Pegasi b Fellow though the Heising-Simons Foundation. AV's work was performed under contract with the California Institute of Technology (Caltech)/Jet Propulsion Laboratory (JPL) funded by NASA through the Sagan Fellowship Program executed by the NASA Exoplanet Science Institute. This material is based upon work supported by the National Science Foundation Graduate Research Fellowship Program under Grant No. DGE-1650116 to PCT. This work was supported by the \tess{} Guest Investigator program (Grant 80NSSC18K1586, awarded to ACR). This paper includes data collected by the \tess{} mission, which are publicly available from the Mikulski Archive for Space Telescopes (MAST).Funding for the \tess{} mission is provided by NASA’s Science Mission directorate. This work is based [in part] on observations made with the Spitzer Space Telescope, which is operated by the Jet Propulsion Laboratory, California Institute of Technology under a contract with NASA. Support for this work was provided by NASA through an award issued by JPL/Caltech. 
This work makes use of observations from the LCOGT network. This work is based in part on observations obtained at the Southern Astrophysical Research (SOAR) telescope, which is a joint project of the Ministerio da Ci\~encia, Tecnologia, Inova\c c\~oes e Comunica\c c\~oes (MCTIC) do Brasil, the U.S. National Optical Astronomy Observatory (NOAO), the University of North Carolina at Chapel Hill (UNC), and Michigan State University (MSU). Some of the observations reported in this paper were obtained with the Southern African Large Telescope (SALT) through Dartmouth College.
Some of the data presented in this paper were obtained from the Mikulski Archive for Space Telescopes (MAST). STScI is operated by the Association of Universities for Research in Astronomy, Inc., under NASA contract NAS5-26555. The authors acknowledge the Texas Advanced Computing Center (TACC) at The University of Texas at Austin for providing HPC resources that have contributed to the research results reported within this paper\footnote{http://www.tacc.utexas.edu}.  This work has made use of data from the European Space Agency (ESA) mission \emph{Gaia} \footnote{https://www.cosmos.esa.int/gaia}, processed by the \emph{Gaia} Data Processing and Analysis Consortium (DPAC)\footnote{https://www.cosmos.esa.int/web/gaia/dpac/consortium}. Funding for the DPAC has been provided by national institutions, in particular the institutions participating in the \emph{Gaia} Multilateral Agreement. The Gaia archive website is https://archives.esac.esa.int/gaia. Based on observations collected at the European Southern Observatory under ESO programme 072.D-0021(B). This research has made use of the VizieR catalogue access tool, CDS, Strasbourg, France. The original description of the VizieR service was published in A\&AS 143, 23. This research has made use of NASA's Astrophysics Data System Bibliographic Services. We acknowledge the use of public \tess{} Alert data from pipelines at the \tess{} Science Office and at the \tess{} Science Processing Operations Center. Resources supporting this work were provided by the NASA High-End Computing (HEC) Program through the NASA Advanced Supercomputing (NAS) Division at Ames Research Center for the production of the SPOC data products.

\vspace{5mm}
\facilities{\emph{TESS}, SALT (HRS), SOAR (Goodman), WASP, \emph{Spitzer}, lightcurve (NRES), CDS, MAST, Simbad}
\software{{\tt Astropy} \citep{astropy13,astropy18}, {\tt emcee} \citep{Foreman-MackeyEmcee2013}, {\tt celerite} \citep{Foreman-MackeyFast2017} {\tt VESPA} \citep{Morton2015}, {\tt SPOC pipeline} \citep{Jenkins:2015,Jenkins:2016}, {\tt misttborb} \citep{JohnsonK22602018}, \tt {BATMAN} \citep{KreidbergBatman2015}, \tt{POET} \citep{StevensonTransit2012,Campo_2011}}

\bibliography{ref_youngplanets}

\begin{deluxetable*}{ccc} 
\tabletypesize{\scriptsize} 
\tablewidth{0pt} 
\tablecaption{Transit Fit Parameters for HIP~67522\,b. \label{tab:transfit}} 
\tablehead{\colhead{Parameter} & \colhead{TESS only} & \colhead{TESS + IRAC CH2}} 
\startdata 
\multicolumn{3}{c}{Fit Parameters}\\ 
\hline 
$T_0$ (BJD)                    & 2458604.02358$^{+0.00041}_{-0.00041}$              & 2458604.02371$^{+0.00036}_{-0.00037}$             \\ 
$P$ (days)                     & 6.95993$^{+0.00034}_{-0.00034}$                    & 6.959503$^{+0.000016}_{-0.000015}$                \\ 
$R_P/R_\star$                  & 0.0645$^{+0.0015}_{-0.0015}$                       & 0.0667$^{+0.0012}_{-0.0011}$                      \\ 
$b$                            & 0.15$^{+0.13}_{-0.1}$                              & 0.134$^{+0.106}_{-0.092}$                         \\ 
$\rho_\star (\rho_\odot)$      & 0.455$^{+0.052}_{-0.049}$                          & 0.448$^{+0.055}_{-0.049}$                         \\ 
$\sqrt{e}\sin{\omega}$         & -0.055$^{+0.103}_{-0.084}$                         & -0.053$^{+0.106}_{-0.084}$                        \\ 
$\sqrt{e}\cos{\omega}$         & 0.00$^{+0.34}_{-0.34}$                             & 0.00$^{+0.34}_{-0.35}$                            \\ 
$q_{1,1}$                      & 0.157$^{+0.099}_{-0.073}$                          & 0.241$^{+0.087}_{-0.068}$                         \\ 
$q_{1,2}$                      & 0.20$^{+0.16}_{-0.13}$                             & 0.22$^{+0.15}_{-0.13}$                            \\ 
$q_{1,2}$                      & ...                                                & 0.014$^{+0.024}_{-0.010}$                         \\ 
$q_{2,2}$                      & ...                                                & 0.160$^{+0.101}_{-0.089}$                         \\ 
\hline 
\multicolumn{3}{c}{GP Parameters}\\ 
\hline 
$\log A_1$                     & -10.45$^{+0.47}_{-0.28}$                           & -10.44$^{+0.37}_{-0.26}$                          \\ 
$\log Q_1$                     & 4.68$^{+0.79}_{-0.74}$                             & 4.59$^{+0.75}_{-0.72}$                            \\ 
$A_2/A_1$                      & 0.9916$^{+0.0081}_{-0.2544}$                       & 0.9905$^{+0.0092}_{-0.2068}$                      \\ 
$\log Q_2$                     & 2.02$^{+0.32}_{-0.25}$                             & 2.00$^{+0.28}_{-0.24}$                            \\ 
$P_{Rot}$ (days)               & 1.422$^{+0.014}_{-0.016}$                          & 1.422$^{+0.014}_{-0.016}$                         \\ 
$\log \sigma$                  & -15.7$^{+3.1}_{-2.9}$                              & -15.3$^{+2.9}_{-3.1}$                             \\ 
\hline 
\multicolumn{3}{c}{Derived Parameters}\\ 
\hline 
$a/R_\star$                    & 11.73$^{+0.24}_{-0.28}$                            & 11.70$^{+0.24}_{-0.27}$                           \\ 
$i$ (deg)                      & 89.27$^{+0.51}_{-0.67}$                            & 89.34$^{+0.45}_{-0.54}$                           \\ 
$t_{14}$ (hours)               & 4.786$^{+0.032}_{-0.030}$                          & 4.822$^{+0.021}_{-0.019}$                         \\ 
$R_P (R_\oplus)$               & 9.72$^{+0.48}_{-0.47}$                             & 10.07$^{+0.47}_{-0.47}$                           \\ 
$e$                            & 0.061$^{+0.172}_{-0.047}$                          & 0.059$^{+0.193}_{-0.046}$                         \\ 
$\omega$ (deg)                 & -21$^{+86}_{-145}$                                 & -17$^{+92}_{-149}$                                \\ 
$T_{eq} (K)$                   & 1173$^{+21}_{-20}$                                 & 1174$^{+21}_{-20}$                                \\ 
$u_{1,1}$                      & 0.148$^{+0.111}_{-0.093}$                          & 0.22$^{+0.12}_{-0.12}$                            \\ 
$u_{1,2}$                      & 0.23$^{+0.16}_{-0.14}$                             & 0.27$^{+0.15}_{-0.15}$                            \\ 
$u_{1,2}$                      & ...                                                & 0.033$^{+0.037}_{-0.021}$                         \\ 
$u_{2,2}$                      & ...                                                & 0.076$^{+0.061}_{-0.043}$                         \\ 
\enddata
\tablenotetext{a}{Equilibrium temperature T$_\mathrm{eq}$ was calculated assuming zero albedo}
\end{deluxetable*} 

\begin{deluxetable*}{cc} 
\tabletypesize{\footnotesize} 
\tablewidth{0pt} 
\tablecaption{Transit Fit Parameters for HIP~67522\,c. \label{tab:transfitc}} 
\tablehead{\colhead{Parameter} & \colhead{} }
\startdata 
$T_0$ (BJD)                    & 2458602.5026$\pm$0.0014                \\ 
$P$ (days)                     & 54$^{+70}_{-24.0}$                                \\ 
$R_P/R_\star$                  & 0.0532$^{+0.0044}_{-0.0041}$                      \\ 
$b$                            & 0.52$^{+0.23}_{-0.33}$                            \\ 
$e$                            & 0.29$^{+0.15}_{-0.14}$                            \\ 
$\omega^a$ (deg)                 & 74$^{+56}_{-53.0}$                                \\ 
$R_P^a (R_\oplus)$               & 8.01$^{+0.75}_{-0.71}$                            \\ 
$t_{14}^a$ (hours)               & 5.707$^{+0.127}_{-0.09}$                          \\ 
$i^A$ (deg)                      & 89.56$^{+0.29}_{-0.42}$                           \\ 
$a/R_\star^a$                    & 56$^{+58}_{-21.0}$                                \\ 
$T_{eq}^b$ (K) & 573$^{+113}_{-163}$ 
\enddata
\tablenotetext{a}{$R_P$, $\omega$, $e$, $t_{14}$, $i$ and $a/R_\star$ were not directly fit, but derived from the fit parameters and stellar parameters from Table \ref{proptab}.}
\tablenotetext{b}{Equilibrium temperature T$_\mathrm{eq}$ was calculated assuming zero albedo.}
\end{deluxetable*}

\end{document}